\documentclass[11pt,a4paper]{article}
\usepackage{jcappub}
\usepackage{bm}
\usepackage{amsmath}
\usepackage[mathscr]{euscript}
\usepackage{verbatim}
\usepackage{ulem}
\def\dd{\mathrm{d}}

\def\mcP{\mathcal{P}}
\def\mcR{\mathcal{R}}
\def\mcS{\mathcal{S}}

\def\Mpl{M_{\rm Pl}}
\def\GeV{{\rm GeV}}

\title{
Mixed Non-Gaussianity from
Axion-Gauge Field Dynamics
}
\author[a,b]{Tomohiro Fujita,}
\author[c]{Ryo Namba,}
\author[d]{and Ippei Obata}
\affiliation[a]{Department of Physics, Kyoto University, Kyoto, 606-8502, Japan}
\affiliation[b]{D\'epartment de Physique Th\'eorique and Center for Astroparticle Physics, \\
Universit\'e de Gen\`eve, Quai E.
Ansermet 24, CH-1211 Gen\`eve 4, Switzerland}
\affiliation[c]{Department of Physics, McGill University, Montr\'{e}al, QC, H3A 2T8, Canada}
\affiliation[d]{Institute for Cosmic Ray Research, The University of Tokyo,
5-1-5 Kashiwa-no-Ha, Kashiwa, Chiba, 277-8582, Japan}
\emailAdd{tomohiro.fujita@etu.unige.ch}
\emailAdd{namba@physics.mcgill.ca}
\emailAdd{obata@icrr.u-tokyo.ac.jp}

\abstract{
We study scalar-tensor-tensor cross correlation $\langle \zeta hh \rangle$ generated by the dynamics of interacting axion and $SU(2)$ gauge fields during inflation. We quantize the quadratic action and solve the linear equations by taking into account mixing terms in a non-perturbative manner. Combining that with the in-in formalism, we compute contributions from cubic interactions to the bispectrum $B_{\zeta hh}$.
We find that the bispectrum is peaked at the folded configuration, which is a unique feature 
encoded by the scalar mixing and localized production of tensor modes. With our parameter choice, the amplitude of the bispectrum is $k^6 B_{\zeta hh} \sim 10^{-16}$.
The unique shape dependence, together with the parity-violating nature, is thus a distinguishing feature to search for in the CMB observables.
}


\usepackage{amsfonts}

\begin{document}

\maketitle

%
%
%
\section{Introduction}
\label{sec:intro}

Detection of primordial gravitational waves (PGWs) is considered to be a smoking gun for the inflationary universe.
Their amplitude is often characterized by the tensor-to-scalar ratio $r$, and a number of experiments aim to pin down the value of $r$ through the B-mode polarizations in the Cosmic Microwave Background (CMB) temperature anisotropy.
While current Planck and BICEP2/Keck Array joint observations have constrained its amplitude as $r \lesssim 0.06$ \cite{Ade:2015lrj}, the observational sensitivity is expected to increase up to $\Delta r = \mathcal{O}(10^{-3})$ in the next decade by upcoming missions including CMB-S4 project \cite{Abazajian:2016yjj} and LiteBIRD satellite \cite{Matsumura:2013aja}.
In the standard prediction, the value of $r$ is directly related to the inflationary energy scale by $E_{\rm inf} \sim 10^{15} \GeV \times ( r / 10^{-3} )^{1/4}$, and thus
if PGWs should be detected by these missions, the energy scale of inflation
would naively be estimated around GUT scales, $\sim 10^{15}\GeV$.
Therefore, in view of studies on fundamental physics, it is extremely important to test the validity of this prediction and to explore viable inflationary mechanisms to provide PGWs.

In the standard inflationary scenario, PGWs are generated by vacuum fluctuations of spacetime amplified due to the quasi-de Sitter expansion of the universe.
The resultant statistical properties are encoded in the tensor spectrum of cosmological perturbations and are (i) nearly scale invariant, (ii) statistically isotropic, (iii) parity symmetric and (iv) almost gaussian.
However, these features are not necessarily true if other sources of gravitational waves do exist in the early universe, such as gauge fields.
In string theory or supergravity, gauge sectors are kinetically or topologically coupled with scalar sectors even if they are neutral under the corresponding gauge group.
Through these couplings a background motion of scalar fields can violate the conformal invariance of gauge fields and amplify gauge quanta during inflation.
Historically their cosmological role has been discussed in the context of primordial magnetogenesis \cite{Ratra:1991bn, Garretson:1992vt, Martin:2007ue, Demozzi:2009fu, Kanno:2009ei, Fujita:2012rb, Ferreira:2013sqa, Fujita:2013pgp, Fujita:2014sna, Obata:2014qba, Fujita:2015iga, Fujita:2016qab, Adshead:2016iae, Caprini:2017vnn}.
Recently, it has been revealed that amplified gauge fields also enhance other fluctuations that are coupled to them and imprint observable signatures in the spectrum of scalar and/or tensor perturbations.
Depending on the dynamics of gauge field production during inflation, the curvature perturbation sourced by produced gauge quanta can be highly non-gaussian \cite{Barnaby:2010vf, Barnaby:2011qe, Barnaby:2012tk, Anber:2012du, Barnaby:2012xt, Linde:2012bt, Ferreira:2014zia, Peloso:2016gqs}, statistically anisotropic \cite{Watanabe:2009ct, Himmetoglu:2009mk, Gumrukcuoglu:2010yc, Watanabe:2010fh, Kanno:2010nr, Watanabe:2010bu, Soda:2012zm, Bartolo:2012sd, Ohashi:2013qba, Ohashi:2013pca, Naruko:2014bxa, Ito:2015sxj, Abolhasani:2015cve, Ito:2017bnn}, and sufficiently large to form primordial black holes after inflation \cite{Linde:2012bt, Garcia-Bellido:2016dkw, Domcke:2017fix, Garcia-Bellido:2017aan, Cheng:2018yyr}.
Also, as sourcing effects on tensor modes, several types of inflationary models with gauge fields have been suggested, predicting scale-dependent, statistically anisotropic, parity-violating, and/or non-gaussian PGWs \cite{Ito:2016aai, Sorbo:2011rz, Cook:2011hg, Mukohyama:2014gba, Choi:2015wva, Namba:2015gja, Domcke:2016bkh, Guzzetti:2016mkm, Obata:2016oym, Obata:2016xcr, Fujita:2017jwq, Ozsoy:2017blg, Fujita:2018zbr}.
These significant deviations from the conventional vacuum modes are potentially testable with the correlations of CMB temperature and polarization anisotropies \cite{Saito:2007kt, Bartolo:2014hwa, Bartolo:2015dga, Bartolo:2017sbu, Thorne:2017jft,Hiramatsu:2018vfw}, laser interferometers \cite{Seto:2006hf, Seto:2006dz, Seto:2008sr, Bartolo:2018qqn} or the measurement of pulsar timing arrays \cite{Kato:2015bye}.
 
In particular, $SU(2)$ gauge fields coupled to an axionic field are known as an interesting alternative PGW source.
Chromo-natural inflation was proposed as the first  axion-$SU(2)$ model for inflation \cite{Adshead:2012kp}, where a large axion-gauge coupling allows the vacuum expectation value (vev) of $SU(2)$ gauge fields to support an isotropic inflationary attractor.
Its background solution is realized with broad parameter region which includes 
seemingly different inflationary models such as gauge-flation and non-canonical single field inflation \cite{Maleknejad:2011jw, SheikhJabbari:2012qf, Adshead:2012qe, Dimastrogiovanni:2012st, Maleknejad:2012fw, Maleknejad:2012dt}, and the background isotropy is stable against small anisotropies of Bianchi I type \cite{Maleknejad:2013npa}.
Although the original scenario is excluded from CMB data \cite{Adshead:2013nka}, extended models have been suggested where additional fields can resolve the observational conflict \cite{Obata:2014loa, Obata:2016tmo, Maleknejad:2016qjz, Dimastrogiovanni:2016fuu, Adshead:2016omu, DallAgata:2018ybl, Domcke:2018rvv}.
Intriguingly, in axion-$SU(2)$ models, the rotationally symmetric background configuration enforces fluctuations of $SU(2)$ gauge fields to have components of scalar and tensor types that interact with density perturbations and gravitational waves, respectively, at the linearized level \cite{Dimastrogiovanni:2012ew, Adshead:2013qp, Namba:2013kia}.
Since one polarization mode of tensor components of gauge fields experiences a tachyonic instability around horizon crossing in this class of models, an exponential enhancement of parity-violating gravitational waves are generated.
Remarkably, the linear interactions of perturbations allow for a parameter region where the tensor components are amplified while the scalar ones are not, which enables to provide sizable amount of chiral gravitational waves consistent with CMB data.
 
In addition to two-point functions, three-point correlations sourced by $SU(2)$ gauge fields are also important observables in this scenario.
Recently, non-linear analyses of this type of models have been explored regarding tensor-tensor-tensor non-gaussianity~\cite{Agrawal:2017awz, Agrawal:2018mrg}, scalar-tensor-tensor mixed non-gaussianity~\cite{Dimastrogiovanni:2018xnn}, and the one-loop contribution to the curvature power spectrum~\cite{Dimastrogiovanni:2018xnn,Papageorgiou:2018rfx}.
Along with such a growing interest, in this paper we calculate a three-point cross-correlation function that is a mixed non-gaussianity between scalar and tensor sectors in the framework of axion-$SU(2)$ model proposed in \cite{Dimastrogiovanni:2016fuu}, where an $SU(2)$ gauge field is coupled to a spectator axion field.
The mechanism of generating such a correlation in this model is multi-fold: first, one of the tensor components of gauge field perturbations are copiously produced by the transient tachyonic instability described in the previous paragraph. The metric tensor modes $h$ (gravitational waves) inherit the effect of this production due to linear mixings. The curvature perturbation $\zeta$ on the other hand gravitationally interacts with scalar modes in the axion-$SU(2)$ sector, which have direct three-point interactions with the gauge-field tensor modes. This way, $\zeta$ and $h$ correlate to induce the scalar-tensor-tensor $\zeta hh$ mixed non-gaussianity, mediated by the gauge-field tensor mode.
As a first step, we focus on the gravitational interaction between $\zeta$ and the axion-field perturbation in the scalar sector.

While a similar cross-correlation has been recently discussed in \cite{Dimastrogiovanni:2018xnn}, we introduce a new calculation approach, in which the mixing effect between the axion and $SU(2)$ fields in the quadratic action is fully taken into account and is not disregarded as a higher-order contribution.
More precisely, we apply a non-perturbative formalism to quantize a coupled system \cite{Nilles:2001fg} and include linear mixing terms in the calculation of the mixed scalar-tensor-tensor non-gaussianity, by employing the in-in formalism \cite{Weinberg:2005vy}.
In order to correctly derive three-point vertices in the interaction Hamiltonian, we find that not only the axion perturbation but also the scalar components of the $SU(2)$ gauge field are relevant, which have been neglected in the previous works \cite{Dimastrogiovanni:2018xnn, Papageorgiou:2018rfx}.
We show that the resultant spectrum is a folded-shape and its non-linearity parameter can be $\mathcal{O}(1)$ with our fiducial parameter choice. While a careful analysis on the signal detectability needs to await future studies, we expect that this novel feature should serve as a distinct signature of the present mechanism, confronted with the future CMB measurements.

This paper is organized as follows.
In Sec.~\ref{sec:overview},  we briefly summarize the setup of our model and then explain our approach to calculate the mixed non-gaussianity. In Sec.~\ref{sec:quadratic}, we show the quadratic action for the scalar and tensor sectors and review how to quantize the system in the initial vacuum. We then derive the cubic interaction Hamiltonian in Sec.~\ref{sec:cubic}. Here we also describe our target observable and obtain its formal expression. We show our results in Sec.~\ref{sec:results}. Sec.~\ref{sec:summary} is devoted for the summary and discussion of our result. We show explicit expressions for shape functions in Appendix \ref{appen:explicit} and compare our approach and result to previous works in Appendix \ref{appen:comparison}.

\section{Model and Our Approach}
\label{sec:overview}

\subsection{The model setup}
\label{subsec:model}

Here we briefly describe our model, while the readers are referred to Ref.~\cite{Dimastrogiovanni:2016fuu} for more detailed discussions.
We consider the following matter action under the general relativity
\begin{equation}
S=\int \dd^4x\sqrt{-g}\left[-\frac{1}{2}\left(\partial\phi\right)^{2}-V(\phi)-\frac{1}{2}\left(\partial\chi\right)^{2}-W(\chi)-\frac{1}{4}F_{\mu\nu}^{a}F^{a, \mu\nu}+\frac{\lambda\chi}{4f}F_{\mu\nu}^{a}\tilde{F}^{a, \mu\nu}\right]\,,
\end{equation}
where $\phi$ is an inflaton with potential $V(\phi)$,
$\chi$ is a pseudo-scalar field (axion) with potential $W(\chi)$, $F_{\mu\nu}^{a}\equiv\partial_{\mu}A^{a}_{\nu}-\partial_{\nu}A^{a}_{\mu}-g \epsilon^{abc} A^{b}_{\mu}A_{\nu}^{c}$ is the field strength of a $SU(2)$ gauge field $A_{\mu}^{a}$ and $\tilde{F}^{a, \mu\nu}\equiv\epsilon^{\mu\nu\rho\sigma}F^{a}_{\rho\sigma}/(2\sqrt{-g})$ is its dual with $\epsilon^{\mu\nu\rho\sigma}$ a flat-spacetime totally anti-symmetric symbol of the choice $\epsilon^{0123} = +1$. 
The constants $\lambda$ and $f$ are dimensionless and dimensionful parameters of the model, respectively.
Although we do not specify any concrete model of $V(\phi)$, the inflaton is assumed to cause a quasi-de Sitter expansion $a(t)\propto e^{H t}$ with a nearly constant Hubble parameter $H$ during inflation and to produce the observed amplitude of the curvature perturbation $\zeta$ through the relation,
\begin{equation}
\zeta=-H\frac{\delta\varphi}{\dot{\phi}_0} \; ,
\label{zeta-def}
\end{equation}
in the spatially flat gauge,
where the inflaton $\phi(t,\bm x)$ is split into the background part $\phi_0(t)$ and the perturbation $\delta\varphi(t,\bm x)$, and dot denotes the cosmic time derivative $\dot{}\equiv \partial_t$\,.
We assume that the coupling between the axion and the gauge fields is sufficiently strong and the background value of the gauge field is not negligibly small compared to that of the axion, while the axion-gauge field system is still a spectator sector in that their energy densities do not alter the background expansion driven by the inflaton $\phi_0$.
In that case, the background fields have an attractor solution where the $SU(2)$ gauge fields take an isotropic configuration,
\begin{equation}
A_0^a(t)=0,
\quad
A_i^a(t)=\delta^a_i a(t)Q(t),
\label{A configuration}
\end{equation}
which is compatible with the Friedmann-Lema\^{i}tre-Robertson-Walker (FLRW) metric.
It is useful to introduce two dimensionless quantities,
\begin{equation}
m_Q\equiv \frac{g Q}{H},
\qquad
\Lambda\equiv \frac{\lambda Q}{f}.
\end{equation}
With them, our assumptions of a strong axion-gauge coupling and a significant vev of the gauge fields are quantified as $\Lambda\gg 1$ and $m_Q\gtrsim 1$, respectively. 
In this regime, one can show that the background attractor solution reads
\begin{equation}
\xi\equiv \frac{\lambda \dot{\chi}_0}{2fH}\simeq m_Q+m_Q^{-1},
\qquad
m_Q\simeq \left(\frac{-g^2 f\,W_\chi(\chi_0)}{3\lambda H^4}\right)^{\frac{1}{3}},
\label{attractor}
\end{equation}
where $\chi_0(t)$ is the background part of the axion field, and subscript $\chi$ on $W$ denotes derivative with respect to $\chi$.
Furthermore, the kinetic part of the energy fraction of the background gauge field $\epsilon_E\equiv (\dot{Q}+HQ)^2/(\Mpl^2H^2)$ and its self-interaction part $\epsilon_B\equiv g^2 Q^4/(\Mpl^2H^2)$
satisfy the simple relation, $\epsilon_B\simeq m_Q^2\epsilon_E$.
Throughout this paper, we work in this attractor regime.%
\footnote{We use $m_Q\simeq 3.5$ and $\Lambda\simeq 160$ in our main calculation in Sec.~\ref{sec:results}.}

Around the above well-behaved background, we introduce perturbations as
\begin{equation}
\chi=\chi_0+\delta\chi, 
\quad
A^a_0=a\partial_aY
\quad
A^a_i=a \delta^a_i (Q+\delta Q)+a\epsilon^{iab}\partial_bU+aT_{ai},
\quad
g_{ij}=a^2(\delta_{ij}+h_{ij}),
\label{pert_decomp}
\end{equation}
where $\delta \chi,Y, \delta Q$ and $U$ are scalar perturbations while $T_{ai}$ and $h_{ij}$ are tensor perturbations. We suppress the vector perturbations as they are irrelevant for our target observable in this paper. The $SU(2)$ gauge freedom is already fixed in the expression \eqref{pert_decomp}, and $Y$ is a non-dynamical variable that can be expressed in terms of dynamical degrees of freedom.
Note that the $SU(2)$ indices and the spacial indices are treated identical under the attractor configuration eq.~\eqref{A configuration} \cite{Maleknejad:2011jw}.%
\footnote{The spatial component of the gauge field vector potential $\bm{A}_i \equiv A^a_i \tau^a$, where $\tau^a$ is the generator of $SU(2)$, transforms under the global part of $SU(2)$ in its adjoint representation, which is isomorphic to $SO(3)$. The vector configuration \eqref{A configuration} precisely enforces to identify this (global) $SO(3)$ with the background spatial rotation.}
As we will see in Sections \ref{sec:quadratic} and \ref{sec:cubic}, these perturbations have interactions in quadratic and cubic actions in this model, which lead to the mixed non-gaussianity.

\subsection{New calculation approach}
\label{subsec:outline}

In this paper we compute the scalar-tensor-tensor non-gaussianity $\langle\zeta hh\rangle$   by using a new calculation approach.
As we discuss in Sec.~\ref{sec:cubic}, the relation between $\zeta$ and $\delta\chi$ is classical and linear.
Hence it is essential to evaluate $\langle \delta\chi hh\rangle$ in a quantum mechanical way to obtain $\langle \zeta hh\rangle$ in this model, which is concretely shown in Sec.~\ref{sec:cubic}.
Here we explain the reason why we introduce a new approach to compute $\langle \delta\chi hh\rangle$ and outline its calculation scheme.

\begin{figure}[tbp]
  \begin{center}
  \includegraphics[width=90mm]{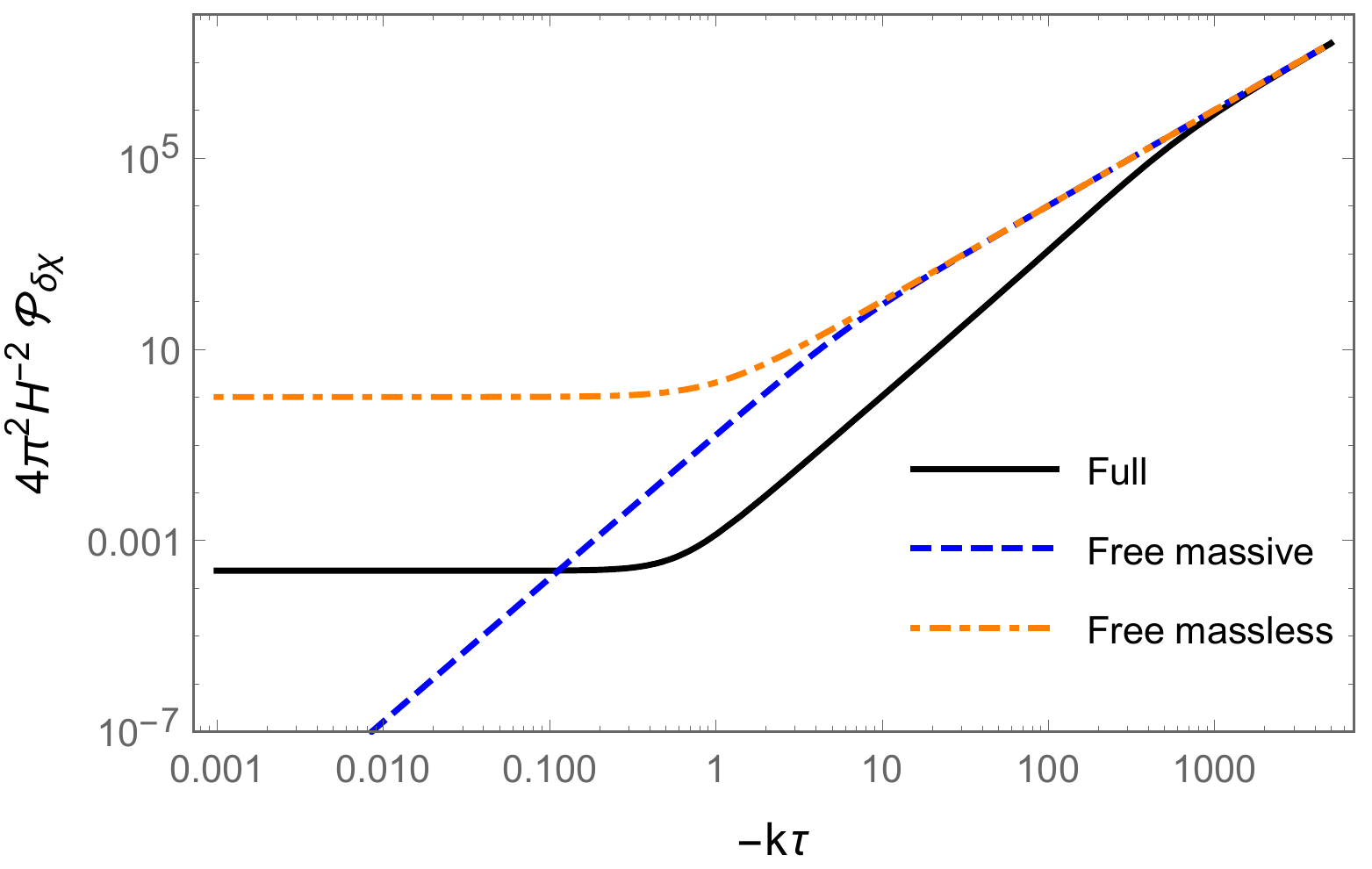}
  \end{center}
  \caption
 {The power spectra of $\delta\chi$ normalized by $(H/2\pi)^2$ which are computed in different ways are compared. The black solid line shows the result of the NPS method where the mixing effects between the scalar perturbations are fully taken into account. 
The other two lines neglect the mixing, and $\delta\chi$ has no mass (orange dot-dashed line) and has a mass $m_\chi^2\approx 40H^2$ (blue dashed line).
The significant deviations imply that one should not regard the mixing effects as a small correction.}
 \label{Pchi-comparison}
\end{figure}

Since $\langle \delta\chi hh\rangle$ is a cross-correlation, some interactions between fields must be involved in the calculation. In our model, we have relevant interaction terms in the quadratic action and cubic action, whose explicit expressions are shown in Sections \ref{sec:quadratic} and \ref{sec:cubic}, respectively.
We call the former  mixing terms (e.g. $\delta \chi \delta Q$), while the latter are called 3-point vertex terms (e.g. $\delta\chi T_{ij} T_{ij}$, and recall that $T$ has linear mixing with $h$). The importance of the mixing effects is illustrated in Fig.~\ref{Pchi-comparison}.
One can see the behavior of $\delta\chi$ is drastically changed by the mixing.
Therefore, although one may obtain a non-zero value of $\langle \delta\chi hh\rangle$ by considering only the 3-point vertices and ignoring the mixing, it is indispensable to take into account the mixing effects in order to properly
evaluate $\langle \delta\chi hh\rangle$.

%
\begin{figure}[tbp]
    \hspace{7mm}
  \includegraphics[width=53mm]{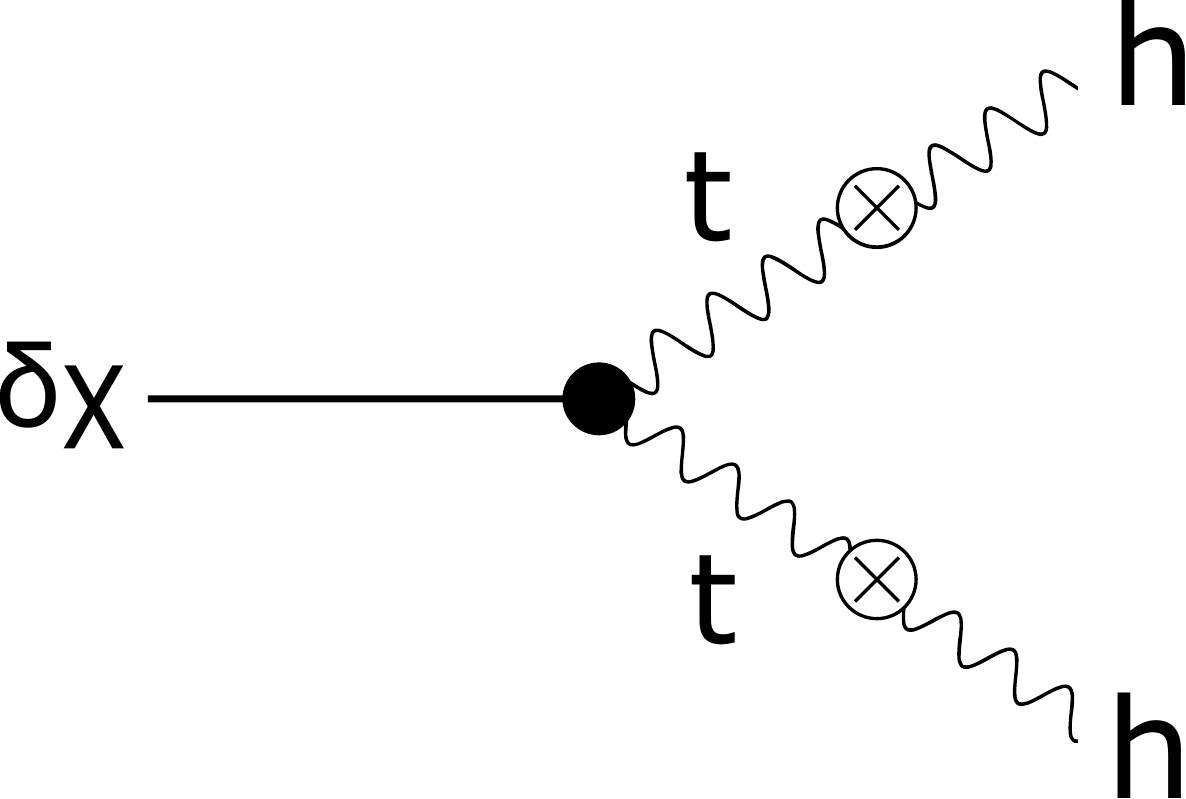}
  \hspace{20mm}
  \includegraphics[width=60mm]{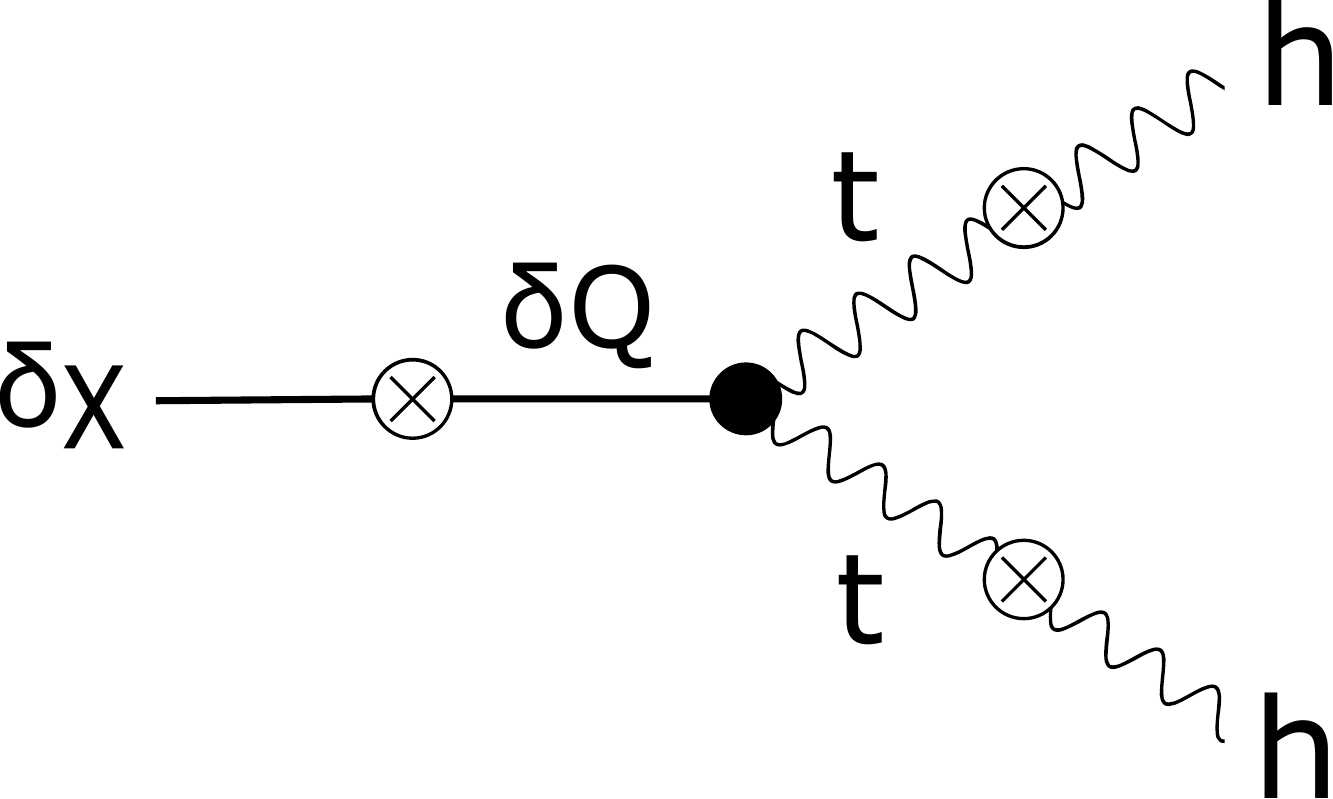}
  \caption
 {These diagrams schematically show the two different channels to produce 
 $\langle\delta\chi h h\rangle$ which are contributed by the vertices
 $\delta\chi TT$ (left) and $\delta Q TT$ (right).
 Black circle and crossed circle denote the 3-point vertex and
 the mixing effect, respectively.
 If the mixing between $\delta\chi$ and $\delta Q$ is treated as a perturbation,
 the right diagram is considered as a higher-order contribution and should be suppressed compared to the left one. Nevertheless, our calculation shows the contributions of the two channels are comparable.}
\label{diagram}
\end{figure}

%
One may think to put both the mixing and vertex terms into the 
interaction Hamiltonian in the in-in formalism all together and perturbatively calculate their contributions to $\langle \delta\chi hh\rangle$.
In that case, one would expect that the contribution from the diagram in the right panel of Fig.~\ref{diagram} is suppressed compared to the left panel contribution, because an extra mixing effect is involved in the right panel. 
Nevertheless, we find that this naive perturbative counting fails and the two contributions are actually comparable.
It suggests that the mixing effects should be taken into account in a non-perturbative manner.

Why is the mixing effects between the scalar perturbations so significant that a perturbative treatment is invalidated in the present model?
To understand the reason, it is useful to consider the background dynamics.
In the previous subsection, we have assumed that the coupling between the axion and the gauge fields is strong enough to achieve the slow-roll regime.
The background axion $\chi_0$, which is pushed by its own potential force, acquires a (non-Hubble) friction from the gauge field $Q_0$.
At the same time, as the backreaction from $\chi_0$ to $Q_0$, the gauge field in turn sustains its background value $Q_0$ by gaining the kinetic energy of $\chi_0$.
If $Q_0$ becomes too big (small), the friction to $\chi_0$ increases (decreases) and the energy transfer from $\dot{\chi}_0$ to $Q_0$ 
diminishes (enlarges). As a result of this continuous mutual feedback system, $\chi_0$ and $Q_0$ end up balancing each other out and keep the slow-roll regime.
When it comes to the perturbations, the scalar degrees of freedom, $\delta\chi, \delta Q$ and $U$, must have similar properties to the background, because their behaviors should be converged to those of the background fields in the long wavelength limit. 
Their leading interactions (or feedback) are represented by the mixing terms. If one employed the in-in formalism or equivalently the Green function method and computed the mixing effects only perturbatively, the interaction of one way (e.g.~$\delta Q$ slows down $\delta\chi$) and its subsequent backreaction ($\delta\dot{\chi}$ sources $\delta Q$) would be treated hierarchically and would not be in the same order in perturbations. As a result, one would not correctly reproduce the mutual feedback system.%
\footnote{This argument does not apply to the tensor perturbations. Because of the hierarchy $T^R\gg h^R$ due to a suppressed mixing between them (the right-handed modes ($R$) are the ones amplified with our choice of parameters), one can focus on the source effect from $T^R$ to $h^R$ and its backreaction is negligible to compute $h^R$. Thus, the result of the non-perturbative calculation is well approximated by a perturbative one in the case of the tensor perturbations.}
It is therefore mandatory to treat the mixing in the scalar sector non-perturbatively.
%
%

Fortunately, a non-perturbative quantum formalism to include the mixing terms is known. 
Nilles, Peloso and Sorbo (NPS) have developed the formalism in Ref.~\cite{Nilles:2001fg}, and Dimastrogiovanni and Peloso have applied it to the axion-$SU(2)$ coupled system in the context of Chromo-natural inflation~\cite{Dimastrogiovanni:2012ew}.
With the NPS formalism, one can quantize a coupled quadratic action and solve the corresponding coupled system of linear equations even with arbitrary mixing. 
On the other hand, the in-in formalism is the technique that readily takes care of cubic and higher order interaction terms.
Therefore, the state-of-art calculation approach is the combination of the NPS method and the in-in formalism: We solve the linear equations of motion derived from the quadratic action including all the mixing terms through the NPS formalism. Then, we calculate the contribution of each 3-point vertex to  $\langle \delta\chi hh\rangle$ by using the in-in formalism.
Note that we demonstrate this hybrid approach for the first time in the literature.

\section{Quadratic Action with Non-perturbative Treatment}
\label{sec:quadratic}

In this section, we write down the quadratic actions of the scalar and tensor perturbations, quantize them with the NPS method developed in~\cite{Nilles:2001fg}, and numerically solve their equations of motion (EoM). 
First, we transform the perturbations into Fourier space,
\begin{align}
S_I(\tau,\bm x) &= \int \frac{\dd^3 k}{(2\pi)^{3/2}}\, {\rm e}^{i\bm k \cdot \bm x} \hat S_I(\tau,\bm k),
\\
T_{I, ij}(\tau,\bm x) &= \sum_{\sigma=L,R}\int \frac{\dd^3 k}{(2\pi)^{3/2}}\, {\rm e}^{i\bm k \cdot \bm x}\, e_{ij}^{\sigma}(\hat{\bm k})\, \hat T_I^\sigma(\tau,\bm k),
\end{align}
where subscript $I (=\chi, Q,U, h,T)$ is the label of the perturbations, 
\begin{equation}
\hat S_\chi \equiv \delta\hat\chi, \quad \hat S_Q\equiv \delta \hat Q,\quad
\hat S_U \equiv k \hat U,\quad \hat T_h^\sigma \equiv \hat h^\sigma,\quad \hat T_T^\sigma=\hat T^\sigma,
\label{SITI-def}
\end{equation}
hat denotes operators in Fourier space, $\sigma$ is the label of the tensor polarizations, and $e_{ij}^{L,R}$ are the circular polarization tensors satisfying
$e^{L}_{ij} (-\hat{\bm{k}})= e^{L*}_{ij} (\hat{\bm{k}})=e^{R}_{ij} (\hat{\bm{k}}),\ 
i \epsilon_{ijk} k_i e_{jl}^{L/R}(\hat{\bm{k}})=
\pm k e_{kl}^{L/R}(\hat{\bm{k}})$.%
\footnote{See e.g.~\cite{Agrawal:2018mrg} for details. We report in passing that there is a typo in \cite{Dimastrogiovanni:2016fuu}, and the definition of $L$ and $R$ below eq.~(2.12) in that paper should be inverted, while all the subsequent calculations were done consistently with the same definition as the one in this paper.}
Note that we have switched the time variable to conformal time $\tau$, which is useful to analyze the perturbations.

After integrating out the non-dynamical variable $Y$, we obtain the scalar quadratic action $S_S^{(2)}$ in Fourier space as,
\begin{equation}
S_S^{(2)}=\frac{1}{2}\int \dd\tau\,\dd^3 k
\left[
\hat\Delta'^\dag_I \hat\Delta'_I+ \hat\Delta'^\dag_I K_{IJ} \hat\Delta_J- \hat\Delta^\dag_I K_{IJ} \hat\Delta'_J
- \hat\Delta^\dag_I \Omega^2_{IJ} \hat\Delta_J\right],
\label{scalar action}
\end{equation}
where prime denotes derivative with respect to $\tau$ and the canonical fields $\Delta_I$ are defined as
\begin{equation}
\hat S_I=\mcR_{IJ}\hat\Delta_J ,
\qquad 
\mcR_{IJ}\equiv 
\left(
\begin{array}{ccc}
\displaystyle \frac{1}{a} \;\; & 0 & 0 \\
0 \;\; & \displaystyle \frac{1}{\sqrt{2} \, a} \;\; & 0 \\
0 \;\; & \displaystyle - \frac{m_Q H}{\sqrt{2} \, k} \;\; & 
\displaystyle \frac{{\cal B}}{\sqrt{2} \, k \, a}
\end{array}
\right)
\label{scalar-rotation}
\end{equation}
and the matrices $K_{IJ}$ and $\Omega^2_{IJ}$ are given by
\begin{equation}
\begin{aligned}
K_{IJ} & = \frac{a \Lambda}{\sqrt{2}}
\left(
\begin{array}{ccc}
0 & \displaystyle m_Q H \;\; & \displaystyle \frac{a m_Q^2 H^2}{{\cal B}} \\
- \displaystyle m_Q H & 0 \;\; & 0 \\
\displaystyle - \frac{a m_Q^2 H^2}{{\cal B}} & 0 \;\; & 0
\end{array}
\right)
\; ,
\\ 
\Omega^2_{IJ} & =
\left(
\begin{array}{ccc}
\displaystyle 
k^2 + a^2 W_{\chi\chi} - \frac{a''}{a}
+ \frac{a^2 k^2 m_Q^2 \Lambda^2 H^2}{{\cal B}^2}  \;\;
& \displaystyle 
\frac{a^2 m_Q \Lambda H^2}{\sqrt{2}}
\left( 3 + \frac{2 \, Q'}{a Q H} \right)  \;\;
& \displaystyle \Omega^2_{\chi U} \\
\displaystyle 
& 
k^2 + 2 a^2 m_Q \left( 2 m_Q - \xi \right) H^2 \;\;
& 
\Omega^2_{QU} \\
\displaystyle  & \;\; 
& 
\Omega^2_{UU}
\end{array}
\right)
\; , \\ &
\Omega^2_{\chi U} = \frac{\Lambda}{\sqrt{2} \, {\cal B}} \left[ a^3 m_Q^2 H^3 + \frac{2k^4 +3 a^2 k^2 m_Q^2 H^2 + 4 a^4 m_Q^4 H^4}{{\cal B}^2} \, \frac{\partial_\tau ( a Q )}{aQ} \right] \; ,
\\ &
\Omega^2_{QU} = 2 a \left( m_Q - \xi \right) H {\cal B} \; ,
\\&
\Omega^2_{UU} = k^2 + 2 a^2 m_Q^2 H^2 + \frac{2 a^2 k^2 m_Q \left( m_Q - \xi \right) H^2}{{\cal B}^2} + \frac{6 a^2 k^2 m_Q^2 H^2}{{\cal B}^4} \, \frac{\left[ \partial_\tau (a Q) \right]^2}{a^2 Q^2} \; ,
\end{aligned}
\label{scalar-matrices}
\end{equation}
where ${\cal B} \equiv \sqrt{k^2 + 2a^2 m_Q^2 H^2}$, and $\Omega^2_{IJ}$ is a symmetric matrix.
Note that the off-diagonal components of $K_{IJ}$ and $\Omega^2_{IJ}$ denote the mixing terms among the scalar perturbations.
Also note that the index $U$ on $K_{IJ}$ and $\Omega^2_{IJ}$ in fact corresponds to a linear combination of the original variables $U$ and $\delta Q$.

Following the NPS method, we quantize $\hat\Delta_I$ as
\begin{equation}
\hat{\Delta}_I(\tau,\bm k)=\mathscr{S}_{IJ}(\tau,k)\, \hat{a}_J(\bm k)+
\mathscr{S}_{IJ}^*(\tau,k)\, \hat{a}^\dag_J(-\bm k),
\label{DI-decompose}
\end{equation}
where the creation/annihilation operators satisfy the standard commutation relation
\begin{equation}
\left[\hat{a}_I(\bm k), \hat{a}^\dag_J(\bm k')\right]= \delta_{IJ}
\delta(\bm k-\bm k'),
\qquad
({\rm otherwise})=0.
\label{commutation}
\end{equation}
It should be emphasized that the scalar mode functions $\mathscr{S}_{IJ}(\tau,k)$ have $3\times 3$ components, 
and the $(I,J)$ component represents the part of $I$ field which is induced by the part of $J$ field arising from vacuum fluctuation.
For instance, $\mathscr{S}_{\chi\chi}$ denotes the amplitude of the intrinsic fluctuation of $\delta\chi$, while $\mathscr{S}_{\chi Q}$ denotes the amplitude of $\delta\chi$ which is sourced by the intrinsic $\delta Q$.
Note that $\mathscr{S}_{IJ}$ accompanies the creation/annihilation operator of the $J$ field because $\mathscr{S}_{IJ}$ originates from the vacuum fluctuation of the $J$ field.
This quantization scheme is essential in order to diagonalize the quadratic Hamiltonian in a vigorous manner under the condition that mass matrix $\Omega^2$ in an action of the form \eqref{scalar action} cannot be diagonalized while keeping the kinetic term intact, which is the case for our consideration.

\begin{figure}[tbp]
    \hspace{-2mm}
  \includegraphics[width=74mm]{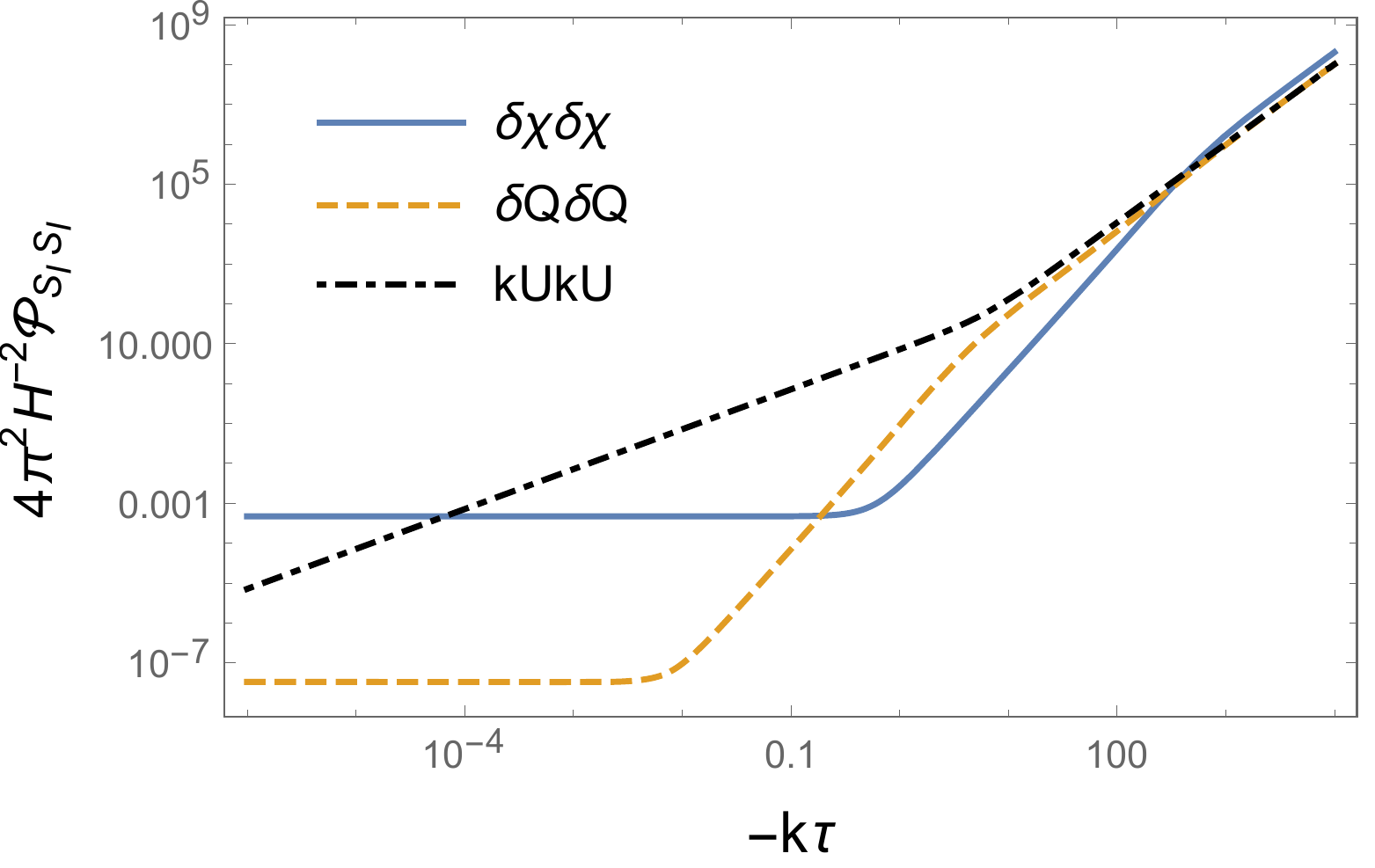}
  \hspace{5mm}
  \includegraphics[width=74mm]{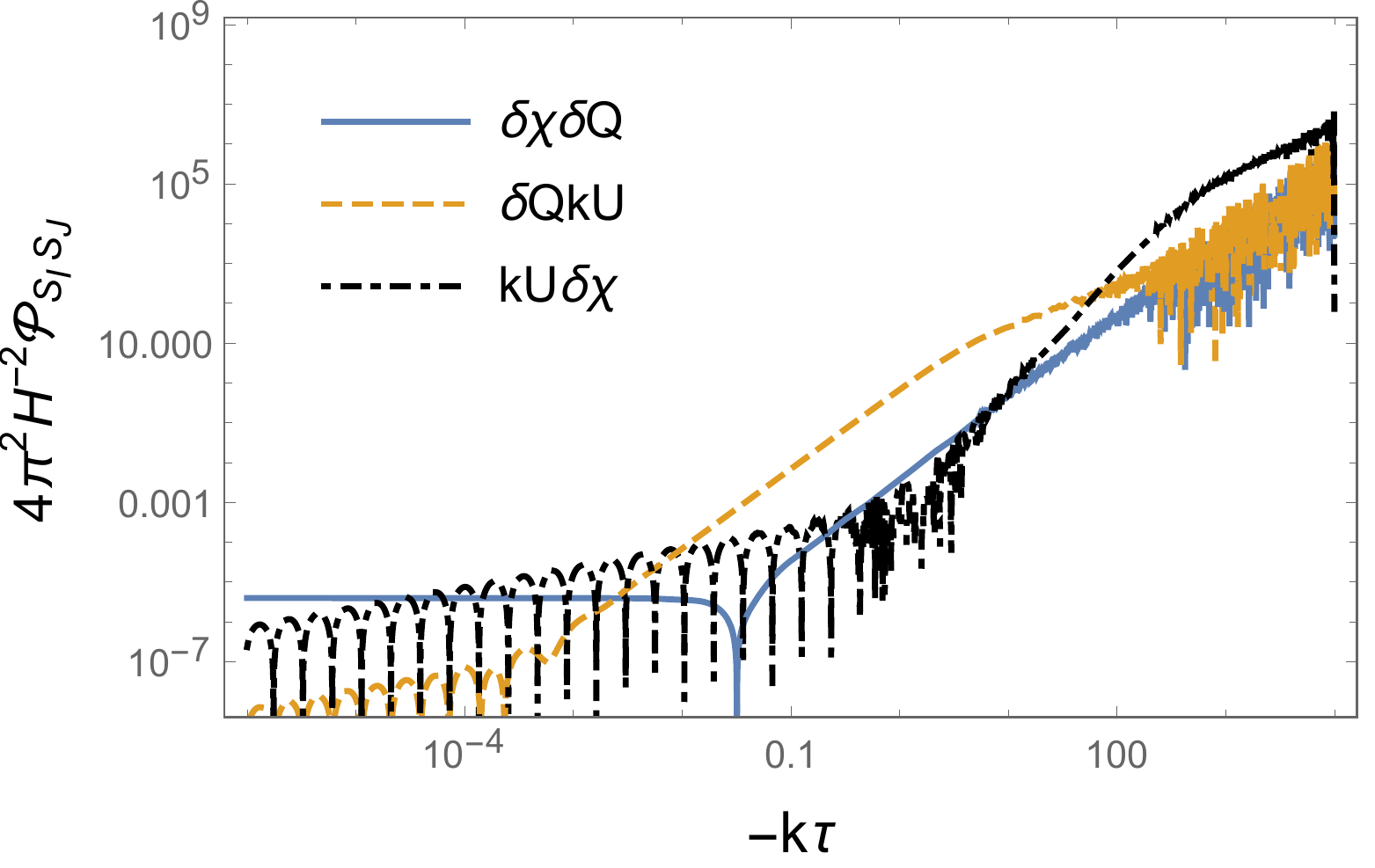}
  \caption
 {{\bf (Left panel)} The auto-power spectra of $\delta \chi$ (blue solid),
 $\delta Q$ (orange dashed) and $kU$ (black dot-dashed) are shown. 
 {\bf (Right panel)} The cross-power spectra between $\delta \chi \delta Q$ (blue solid), $\delta Q k U$ (orange dashed) and $\delta \chi kU$ (black dot-dashed) are plotted. These cross-correlations are non-zero at the linear level by virtue of the mixing terms.  }
 \label{PSplot}
\end{figure}

We numerically solve the EoM for the scalar mode functions $\mathscr{S}_{IJ}(\tau,k)$
\begin{equation}
\partial_\tau^2\mathscr{S}_{IL}+2K_{IJ}\partial_\tau\mathscr{S}_{JL}+(\Omega^2_{IJ} + \partial_\tau K_{IJ})\mathscr{S}_{JL}=0,
\label{scalarEoM}
\end{equation}
with the Bunch-Davies initial condition
\begin{equation}
\lim_{|k\tau|\to\infty}\mathscr{S}_{IJ}=\frac{1}{\sqrt{2k}}\,\delta_{IJ},
\qquad 
\lim_{|k\tau|\to\infty}\partial_\tau \mathscr{S}_{IJ}=-i\sqrt{\frac{k}{2}}\,\delta_{IJ}.
\label{BDvacuum}
\end{equation}
In actual numerical computations, the initial vacuum must be taken at time $\vert k\tau \vert \gg 1$ when adiabatic conditions are satisfied and $\Omega^2_{IJ} \simeq k^2 \delta_{IJ}$ is a good approximation.
The auto- and cross-power spectra of the scalar perturbations are defined by (see e.g.~\cite{Gumrukcuoglu:2010yc})
\begin{equation}
\frac{1}{2}\left\langle \hat{S}_I(\tau,\bm k)\hat{S}_J(\tau,\bm k')
+ \hat{S}_J(\tau,\bm k)\hat{S}_I(\tau,\bm k')\right\rangle\equiv \delta(\bm k+\bm k') \, \frac{2\pi^2}{k^3}\mathcal{P}_{S_I S_J}(\tau,k).
\end{equation}
We show the auto-power spectra $\mcP_{\delta\chi\delta\chi},\mcP_{\delta Q\delta Q}$ and $\mcP_{kUkU}$ in the left panel of Fig.~\ref{PSplot} and the cross-power spectra  $\mcP_{\delta\chi\delta Q},\mcP_{\delta Q kU}$ and $\mcP_{kU\delta\chi}$ in the right panel.
These non-vanishing correlations between $\delta\chi,\delta Q$ and $U$ at the linear order lead to important contributions to the mixed non-gaussianity as we will see in Sec.~\ref{sec:cubic}.
%
%

\begin{figure}[tbp]
    \hspace{-2mm}
  \includegraphics[width=74mm]{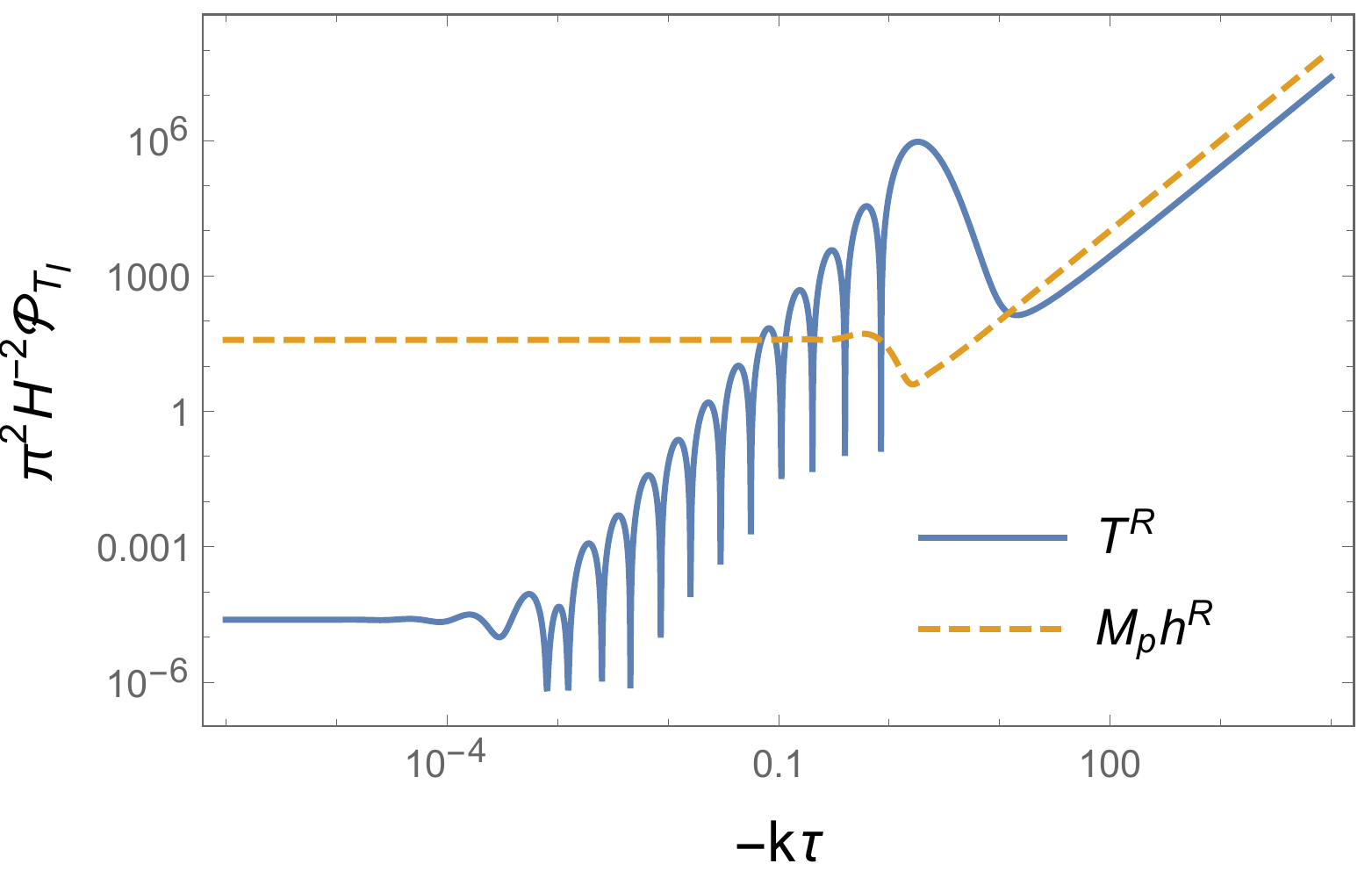}
  \hspace{5mm}
  \includegraphics[width=74mm]{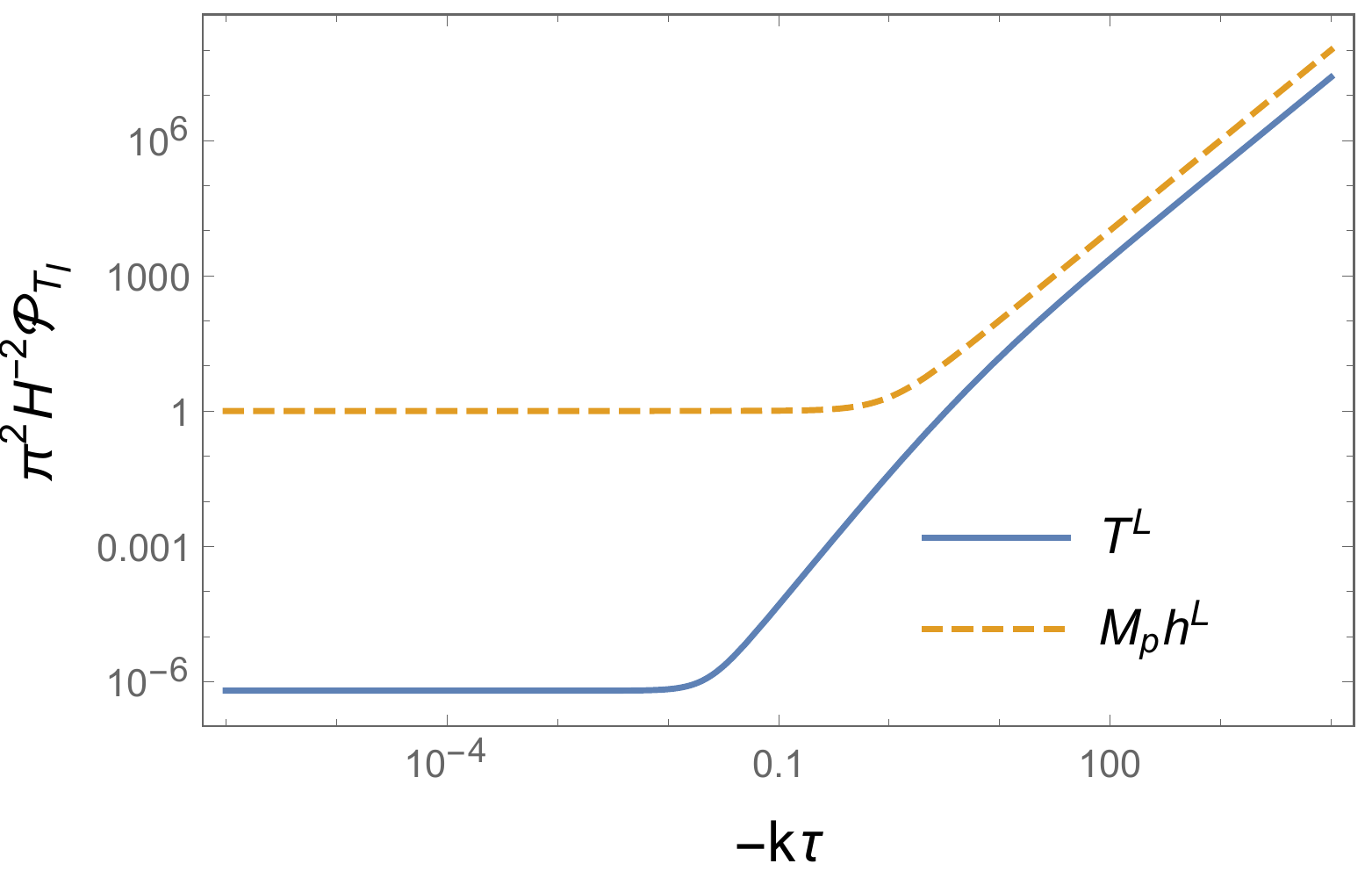}
  \caption
 {The power spectra of the $SU(2)$ tensorial perturbation $T$ (blue solid) and the gravitational waves $h$ multiplied by the reduced Planck mass $\Mpl$ (orange dashed). In the left and right panel, the right- and left-handed polarization modes are shown, respectively. 
With the present parameters, the right-handed $T$ undergoes a tachyonic instability and gets amplified, while the left-handed $T$  exhibits no amplification. Thus, only the right-handed $h$ is significantly sourced.}
 \label{PTplot}
\end{figure}

The tensor perturbations with mixing terms can be quantized and solved in the same way as the scalar ones. Their quadratic action is written in the same form as eq.~\eqref{scalar action} with the replacements
\begin{equation}
\begin{aligned}
\tilde{\mcR}_{IJ} &\equiv 
\left(
\begin{array}{cc}
\displaystyle
\frac{2}{\Mpl \, a} & 0 \\
0 & \displaystyle \frac{1}{a}
\end{array}
\right) \; ,
\qquad
\tilde{K}_{IJ}=
\left(
\begin{array}{cc}
0 & \displaystyle \frac{\partial_\tau ( a Q)}{\Mpl \, a} \\
\displaystyle - \frac{\partial_\tau ( a Q)}{\Mpl \, a} & 0
\end{array}
\right) \; , \\
\tilde{\Omega}^2_{IJ , L/R} & =
\left( 
\begin{array}{cc}
\displaystyle k^2 - \frac{a''}{a} + \frac{2 Q^2}{\Mpl^2} \left[ a^2 m_Q^2 H^2 - \frac{\left[ \partial_\tau \left( a Q \right) \right]^2}{a^2  Q^2} \right] \;\;\; & 
\displaystyle
- \frac{a H Q}{\Mpl} \left[ \pm 2 k m_Q + 2 a m_Q \xi H - \frac{\partial_\tau \left( a Q \right)}{a Q} \right]
\\
\displaystyle
- \frac{a H Q}{\Mpl} \left[ \pm 2 k m_Q + 2 a m_Q \xi H - \frac{\partial_\tau \left( a Q \right)}{a Q} \right] & \displaystyle k^2 \pm 2 k a \left( m_Q + \xi \right) H + 2 a^2 m_Q \xi H^2
\end{array}
\right) ,
\end{aligned}
\label{tensor-matrices}
\end{equation}
where subscript $L/R$ corresponds to left-/right-handed modes, respectively, and to $\pm$ on the right-hand side in the corresponding order.
While the $L$ and $R$ modes are decoupled from each other at linear order, metric ($\hat{h}$) and gauge-field ($\hat{T}$) tensor modes within each sector have linear mixings.
We decompose each sector in terms of creation and annihilation operators in the same manner as for scalar modes \eqref{DI-decompose}, i.e.
\begin{equation}
\hat\Delta^\sigma_I(\tau , \bm{k}) = \mathscr{T}_{IJ}^\sigma(\tau , k) \, \hat{b}_J^\sigma(\bm{k}) + \mathscr{T}_{IJ}^{\sigma \, *}(\tau , k) \, \hat{b}^{\sigma \, \dagger}_J(-\bm{k}) \; ,
\end{equation}
where $\hat\Delta^\sigma_I \equiv (\tilde{\mcR}^{-1})_{IJ} \hat{T}^\sigma_{J}$ are canonically normalized fields for the tensor, and $\hat{b}_J^\sigma$ and $\hat{b}^{\sigma \, \dagger}_J$ have commutation relations of the same form as in \eqref{commutation}.
The matrix of tensor mode functions $\mathscr{T}_{IJ}^\sigma$ has $2\times 2$ components for both polarizations $\sigma=L,R$ and are given the Bunch-Davies initial conditions as in \eqref{BDvacuum}. 
In Fig.~\ref{PTplot}, we show their power spectra defined as $\langle \hat{T}_I^\sigma(\tau,\bm k)\hat{T}_I^\sigma(\tau,\bm k')\rangle\equiv \delta(\bm k+\bm k') 2\pi^2\mcP_{T_I^\sigma}(\tau,k)/k^3$.
As discussed in previous works, the right-handed $SU(2)$ tensor mode undergoes a tachyonic instability around the horizon crossing and is exponentially amplified. The right-handed gravitational waves that are sourced by the amplified $SU(2)$ tensor is substantially enhanced. On the other hand, the left-handed $SU(2)$ tensor mode does not have this instability and hence does not produce an interesting signature.
Henceforth, we concentrate only on the right-handed modes of the tensor perturbations and disregard the left-handed.
%
%

\section{Cubic Action with In-in Formalism}
\label{sec:cubic}

In this and following sections we compute $3$-point correlation functions in this model. Our focus in this paper is cross correlations between scalar ($S$) and tensor ($T$) perturbations. While possible combinations of $3$-point functions are $STT$ and $SST$, the exponential enhancement due to transient tachyonic instability occurs only in the tensor sector. This implies that the number of tensor modes involved in correlation functions counts the order of the exponential amplification.%
\footnote{This argument is not exactly valid in the cases of auto-correlations. Since phases of mode functions freeze once the modes become classical, their Green functions do not experience such enhancements \cite{Ferreira:2015omg, Agrawal:2017awz,Agrawal:2018mrg}. Cross correlations we consider in this paper, on the other hand, do not suffer such cancellation, and the naive power counting described here applies.} 
Therefore, we are only interested in $STT$ correlations.
In the spatially flat gauge $\delta g_{ij} =a^2 \left( \delta_{ij} + h_{ij} \right)$, where $h_{ij}$ is the traceless transverse part of the metric perturbations, $\zeta$ is directly related to the inflaton perturbation $\delta\varphi$ through $\zeta = - H \delta\varphi / \dot\phi_0$, where $\phi_0(t)$ is the vacuum expectation value (vev) of inflaton.%
\footnote{Even if all the energetically subdominant components decay away after or during inflation,  inflaton perturbation $\delta\varphi$ which is induced by them survives and contributes to curvature perturbation $\zeta$.
We conservatively concentrate on it in this paper. 
Some other contributions are discussed in appendix.~\ref{appen:comparison}.}
The inflaton has no direct coupling to the $SU(2)$ gauge field, and thus it receives the effects of gauge field production only through gravitational interactions. The dominant channel among them is the gravitational coupling between $\delta\varphi$ and other scalar modes \cite{Ferreira:2014zia, Namba:2015gja}.
We focus on the coupling with $\delta\chi$, and leave those with the other scalar modes $\delta Q$ and $U$ to future studies. 
The gravitational coupling between $\delta\varphi$ and $\delta\chi$ can in fact be treated perturbatively, and one can show that part of $\delta\hat\varphi(\bm{k})$ due to this coupling, denoted by $\delta_m\hat\varphi(\bm{k})$, is well approximated by $\delta_m\hat\varphi(\bm{k}) = \dot\phi_0 \dot\chi_0 \Delta N_{\chi , k} \delta\hat\chi(\bm{k}) / (\Mpl^2 H^2)$ in terms of the Fourier modes, where $\Delta N_{\chi,k}$ is the number of e-folds after a given mode of $\delta\hat\chi$ crosses the Hubble radius until it starts decaying during or at the end of inflation \cite{Namba:2015gja, Dimastrogiovanni:2016fuu}. Since $\delta\hat\chi$ directly couples to the $SU(2)$ gauge field, $\delta_m\hat\varphi$ feels the gauge field production through the gravitational coupling.
Thus, in order to compute the dominant contribution to $\zeta$ sourced by the production, we consider the part contributing to \eqref{zeta-def},
\begin{equation}
\hat\zeta^{(s)} (\bm{k}) = - \frac{H}{\dot\phi_0} \, \delta_m\hat\varphi(\bm{k})
\simeq - \frac{\dot\chi_0}{\Mpl^2 H} \, \Delta N_{\chi,k} \, \delta\hat\chi (\bm{k}) \; ,
\label{sourced_zeta}
\end{equation}
where superscript $(s)$ denotes sourced part.
The $STT$ $3$-point correlation function in turn yields
\begin{equation}
\langle \hat\zeta(\bm{k}_1) \, \hat h^{\sigma}(\bm{k}_2) \, \hat h^{\sigma'}(\bm{k}_3) \rangle \simeq
- \frac{\dot\chi}{\Mpl^2 H} \, \Delta N_{\chi,k_1} \, 
\langle \delta\hat\chi (\bm{k}_1) \, \hat h^{\sigma}(\bm{k}_2) \, \hat h^{\sigma'}(\bm{k}_3) \rangle \; ,
\label{zetahh-chihh}
\end{equation}
where $\sigma, \sigma'= L,R$ are tensor polarizations.
As argued in Sec.~\ref{subsec:outline}, therefore, the computation of the $STT$ correlation amounts to that of $\langle \delta\chi hh \rangle$.

We employ the in-in formalism, perturbation theory of operator formulation in which correlations of Heisenberg-picture operators are evaluated as expectation values on the ``in'' vacuum \cite{Weinberg:2005vy}.
In computing the $3$-point function as in \eqref{zetahh-chihh}, the leading-order, tree-level, contribution comes from the $STT$ part of cubic interaction Hamiltonian $\delta_3 H_{\rm int}^{STT}$ that correlates with $\delta\chi hh$. 
In deriving $\delta_3 H_{\rm int}^{STT}$, some extra cares need to be taken. First, we need to use constraint equations to solve for the non-dynamical variable $Y$ in favor of dynamical ones $\delta\chi$, $\delta Q$, $U$ and $T_{ij}$. Since $\delta_3 H_{\rm int}^{STT}$ contains terms proportional to $Y S$ and $Y TT$, where $S = \{ \delta\chi , \delta Q , U \}$, solving the constraint equations as $Y = {\cal O}(TT)$ and $Y = {\cal O} (S)$ respectively leads to $STT$ interaction terms.
Secondly, we ignore $Shh$ vertex terms out of $STT$ interaction terms.
The dominant part of right-handed gravitational waves is sourced by the gauge field ``tensor'' mode $T^R$ (strictly speaking $\mathscr{T}_{TT}^R$) due to the mixing. Although $T^R$ is directly coupled to the scalar modes $S$, they have only Planck-suppressed interactions with $h$, which can be treated perturbatively as in \eqref{sourced_zeta}.
Thus, part of $\delta_3 H_{\rm int}^{STT}$ of our interest consists of $\delta\chi TT$, $\delta Q TT$ and $UTT$, where $T$ now stands for the gauge field ``tensor'' modes.

Once these considerations are taken into account, the explicit expression of $\delta_3 H_{\rm int}^{STT}$ is found to be, in Fourier space,%
\footnote{This can be obtained as minus of cubic Lagrangian, and it is known to coincide with cubic Hamiltonian~\cite{Huang:2006eha}. We have explicitly checked the equivalence.}
\begin{equation}
\begin{aligned}
\delta_3 H_{\rm int}^{STT} & = 
\sum_{\sigma , \sigma'} \int \frac{\dd^3k \, \dd^3p \, \dd^3q}{(2\pi)^{3/2}} \, \delta^{(3)} ( \bm{k} + \bm{p} + \bm{q} )
\, e_{ai}^\sigma (\hat{\bm{p}}) \, e_{aj}^{\sigma'} (\hat{\bm{q}})
\\ & \times 
\Bigg[
\frac{\lambda}{f} \,
\epsilon^{ijk} \, i q_k \, \delta\hat\chi_{\bm{k}} \, \frac{\dd (a \hat{T}^\sigma_{\bm{p}})}{\dd\tau} \, a \hat{T}^{\sigma'}_{\bm{q}}
- \frac{g \lambda}{2 f} \, \delta^{ij} \delta\hat\chi_{\bm{k}} \, 
\frac{\dd}{\dd\tau} \left( a^3 Q \, \hat{T}^\sigma_{\bm{p}} \, \hat{T}^{\sigma'}_{\bm{q}} \right)
\\ & \qquad
- \frac{g \lambda \bar\chi}{2f} \, \delta^{ij} \, \frac{\dd}{\dd\tau} \left( a^3 \delta \hat{Q}_{\bm{k}} \, \hat{T}^\sigma_{\bm{p}} \hat{T}^{\sigma'}_{\bm{q}} \right)
- a^3 g \, \epsilon^{ijk} \, i q_k \, \delta \hat{Q}_{\bm{k}} \, \hat{T}^\sigma_{\bm{p}} \, \hat{T}^{\sigma'}_{\bm{q}}
\\ & \qquad
- a^3 g \, k_i k_j \, \hat{U}_{\bm{k}} \, \hat{T}^\sigma_{\bm{p}} \, \hat{T}^{\sigma'}_{\bm{q}}
\\ & \qquad
- \frac{g \epsilon^{ijk} \, i k_k}{k^2 + 2 a^2 g^2 Q^2}
\left(
a^2 \, \frac{g \lambda Q^2}{f} \, \delta\hat\chi _{\bm{k}}
- \frac{\dd}{\dd\tau} \left( a \delta \hat{Q}_{\bm{k}} \right)
+ 2 a g Q \, \frac{\dd}{\dd\tau} \left( a \hat{U}_{\bm{k}} \right)
- 2 a g \, \partial_\tau(aQ) \, \hat{U}_{\bm{k}} \right)
\\ & \qquad\quad \times
\, \frac{\dd (a \hat{T}^\sigma_{\bm{p}})}{\dd\tau} \, a \hat{T}^{\sigma'}_{\bm{q}} \Bigg] \; .
\end{aligned}
\label{HISTT_Fourier}
\end{equation}
where $\hat{T}_{\bm{k}}^\sigma$ are Fourier modes of $SU(2)$ tensor modes with polarization $\sigma$ and momentum $\bm{k}$, and $\epsilon^{ijk}$ is a flat-space totally anti-symmetric symbol.
The last two lines in \eqref{HISTT_Fourier} are originated from the terms of the form $YS$ and $YTT$.
Since only right-handed ($R$) modes of $\hat{T}^\sigma$ are enhanced with our choice of parameters, we neglect all the terms that contain left-handed modes. Then the above expression \eqref{HISTT_Fourier} can be rewritten compactly as
\begin{equation}
\delta_3 H_{\rm int}^{STT}(\tau) = \sum_{I = \chi , Q , U} \int \frac{\dd^3k \, \dd^3p \, \dd^3q}{(2\pi)^{3/2}} \, \delta^{(3)} (\bm{k} + \bm{p} + \bm{q}) \, \mathcal{F}^{(I)}(\bm{k} , \bm{p} , \bm{q} , \tau) \, \hat{S}_{I, \bm{k}}(\tau) \, \hat{T}^R_{\bm{p}}(\tau) \, \hat{T}^R_{\bm{q}}(\tau) \; ,
\label{F-def}
\end{equation}
where operator $\mathcal{F}^{(I)}$ acts on $\hat{S}_I \hat{T}^R \hat{T}^R$. The leading-order contribution to the correlator in \eqref{zetahh-chihh} in the in-in formulation is given by
\begin{equation}
\begin{aligned}
& \langle \delta\hat\chi_{\bm{k}_1} \hat h^{\sigma}_{\bm{k}_2} \hat h^{\sigma'} _{\bm{k}_3} (\tau) \rangle =
i \int^\tau_{-\infty} \dd\tau' \, \left\langle \left[ \delta_3 H_{\rm int}^{STT} (\tau') , \, \delta\hat\chi_{\bm{k}_1}(\tau) \, \hat h^{\sigma}_{\bm{k}_2}(\tau) \, \hat h^{\sigma'}_{\bm{k}_3} (\tau) \right] \right\rangle \; .
\end{aligned}
\label{chihh_general}
\end{equation}
Since only right-handed modes are produced in the tensor sector, the correlation occurs only with right-handed modes, which can easily be shown explicitly.
We define bispectrum $B_{\zeta hh}(k_1 , k_2 , k_3)$ by
\begin{equation}
B_{\zeta h h} (k_1 , k_2 , k_3) \, \delta^{(3)} (\bm{k}_1 + \bm{k}_2 + \bm{k}_3) \equiv \langle \hat\zeta_{\bm{k}_1} \hat h^{R}_{\bm{k}_2} \hat h^{R}_{\bm{k}_3} \rangle \; ,
\label{bispec-def}
\end{equation}
where a general expression is obtained using \eqref{zetahh-chihh} and calculating \eqref{chihh_general}, 
\begin{equation}
\begin{aligned}
B_{\zeta h h} (k_1 , k_2 , k_3) & = \frac{- 2 \Delta N_{\chi,k_1}}{(2\pi)^{3/2}}  \, \frac{\dot\chi}{\Mpl^2 H}
\sum_{I,J,M,N}
\int^\tau_{-\infty} \dd\tau' 
\left[ \mathcal{F}^{(I)}(- \bm{k}_1 , -\bm{k}_2 , -\bm{k}_3 , \tau')
+ \mathcal{F}^{(I)}(- \bm{k}_1 , -\bm{k}_3 , -\bm{k}_2 , \tau') \right]
\\ & \quad \times
{\rm Im} \left[ S_{\chi J , k_1}(\tau) \, R_{hM , k_2}(\tau) \, R_{hN , k_3}(\tau) \,
S_{IJ , k_1}^*(\tau') \, R_{TM , k_2}^*(\tau') \, R_{TN , k_3}^*(\tau') \right] \; ,
\end{aligned}
\label{Bzetahh_general}
\end{equation}
and the explicit expressions of $\mathcal{F}^{(I)}$ will be given in the next section. Here, for notational brevity, we denote $S_{IJ , k} (\tau) \equiv \mcR_{IK}(\tau , k) \mathscr{S}_{KJ}(\tau , k)$ for the scalar and $R_{IJ , k}(\tau) \equiv \tilde{\mcR}_{IK}(\tau , k) \mathscr{T}^R_{KJ}(\tau , k)$ for the right-handed tensor.
During the computation, we have discarded a disconnected piece. It is manifest from this expression that the result is symmetric under interchange of $\bm{k}_2 \leftrightarrow \bm{k}_3$. Note that $B_{\zeta hh}$ has a mass dimension $-6$.
The fact that this bispectrum is originated from $\delta\chi$ is captured by the overall factor $\dot\chi$, and the fact that it is due to the mixing between $\delta\chi$ and $\delta\varphi$ is by $\Delta N_{\chi,k_1}$.
In the next section, we compute $B_{\zeta hh}$ using numerically calculated mode functions $S_{IJ,k}$ and $R_{IJ,k}$, and show our results of its shape and amplitude from the model.

\section{Result of the Mixed Non-gaussianity}
\label{sec:results}

Our goal is to compute the scalar-tensor-tensor correlation $\langle \zeta h h \rangle$ \eqref{zetahh-chihh}, and the general form of its bispectrum $B_{\zeta hh}$ given by \eqref{Bzetahh_general}. In general, bispectra in a scale-invariant system can be characterized by two quantities: {\it shape} and {\it amplitude}. Due to background homogeneity, the three wave vectors in bispectra form a triangle $\bm{k}_1 + \bm{k}_2 + \bm{k}_3 = 0$, as clearly seen in \eqref{bispec-def}, and thus the norms $k_1$, $k_2$ and $k_3$ uniquely determines the {\it shape} of bispectra. Due to background isotropy, the momentum dependence of bispectra can also be reduced to that on $k_1$, $k_2$ and $k_3$.
Moreover, (approximate) scale invariance in the system results in a scaling relation
\begin{equation}
B_{\zeta hh} ( s k_1 , s k_2 , s k_3 ) = s^{-6} B_{\zeta hh} ( k_1 , k_2 , k_3 ) \; ,
\label{Bzhh_scaleinv}
\end{equation}
at the leading order in the slow-roll expansion.
Taking these considerations into account, {\it shape} is often conveniently defined as
\begin{equation}
{\cal S}_{\zeta hh} \equiv {\cal N} k_1^2 k_2^2 k_3^2 B_{\zeta hh} \; ,
\label{shape-def}
\end{equation}
where ${\cal N}$ is an arbitrary normalization factor, and {\it amplitude} by non-linearity parameter
\begin{equation}
f_{\zeta hh}^{\rm NL} \equiv 
\frac{10}{3 \left( 2 \pi \right)^{5/2} {\cal P}_\zeta^2} \, 
\frac{\prod_i k_i^3}{\sum_i k_i^3} \, 
B_{\zeta hh} \; ,
\label{fNL-def}
\end{equation}
where ${\cal P}_\zeta$ is the power spectrum of curvature perturbation. 
The numerical factor in \eqref{fNL-def} is taken in such a way to coincide with the standard definition of non-linearity parameter for scalar auto non-gaussianity \cite{Komatsu:2001rj}, see also \cite{Barnaby:2010vf}.
With these definitions \eqref{shape-def} and \eqref{fNL-def}, ${\cal S}_{\zeta hh}$ and $f_{\zeta hh}^{\rm NL}$ are scaling free, i.e.
\begin{equation}
{\cal S}_{\zeta hh} ( s k_1 , s k_2 , s k_3 ) = s^0 {\cal S}_{\zeta hh} ( k_1 , k_2 , k_3 ) \; , \qquad
f_{\zeta hh}^{\rm NL} ( s k_1 , s k_2 , s k_3 ) = s^0 f_{\zeta hh}^{\rm NL} ( k_1 , k_2 , k_3 ) \; .
\end{equation}
Since our bispectrum is invariant under exchange of tensor-mode momenta, $\bm{k}_2 \leftrightarrow \bm{k}_3$, but not under exchange of $\bm{k}_1$, it is convenient to normalize $k_1$ and $k_2$ by $k_3$, defining
\begin{equation}
x_1 \equiv \frac{k_1}{k_3} \; , \qquad
x_2 \equiv \frac{k_2}{k_3} \; .
\label{x1x2-def}
\end{equation}
Then the $k$-dependences of ${\cal S}_{\zeta hh}$ and $f_{\zeta hh}^{\rm NL}$ are carried only by $x_1$ and $x_2$, ${\cal S}_{\zeta hh} = {\cal S}_{\zeta hh} ( x_1 , x_2)$ and $f_{\zeta hh}^{\rm NL} = f_{\zeta hh}^{\rm NL} ( x_1 , x_2)$.
Having triangular inequalities on $k_1$, $k_2$ and $k_3$, and barring double-counting between $k_2 \leftrightarrow k_3$, the following region of $x_1$ and $x_2$ exhausts all the shapes without redundancy:
\begin{equation}
x_1 + x_2 \ge 1 \; , \qquad
x_2 + 1 \ge x_1 \; , \qquad
x_2 \le 1 \; ,
\end{equation}
where the third triangular inequality is redundant.
\begin{figure}
\centering
\includegraphics[width=0.9\textwidth]{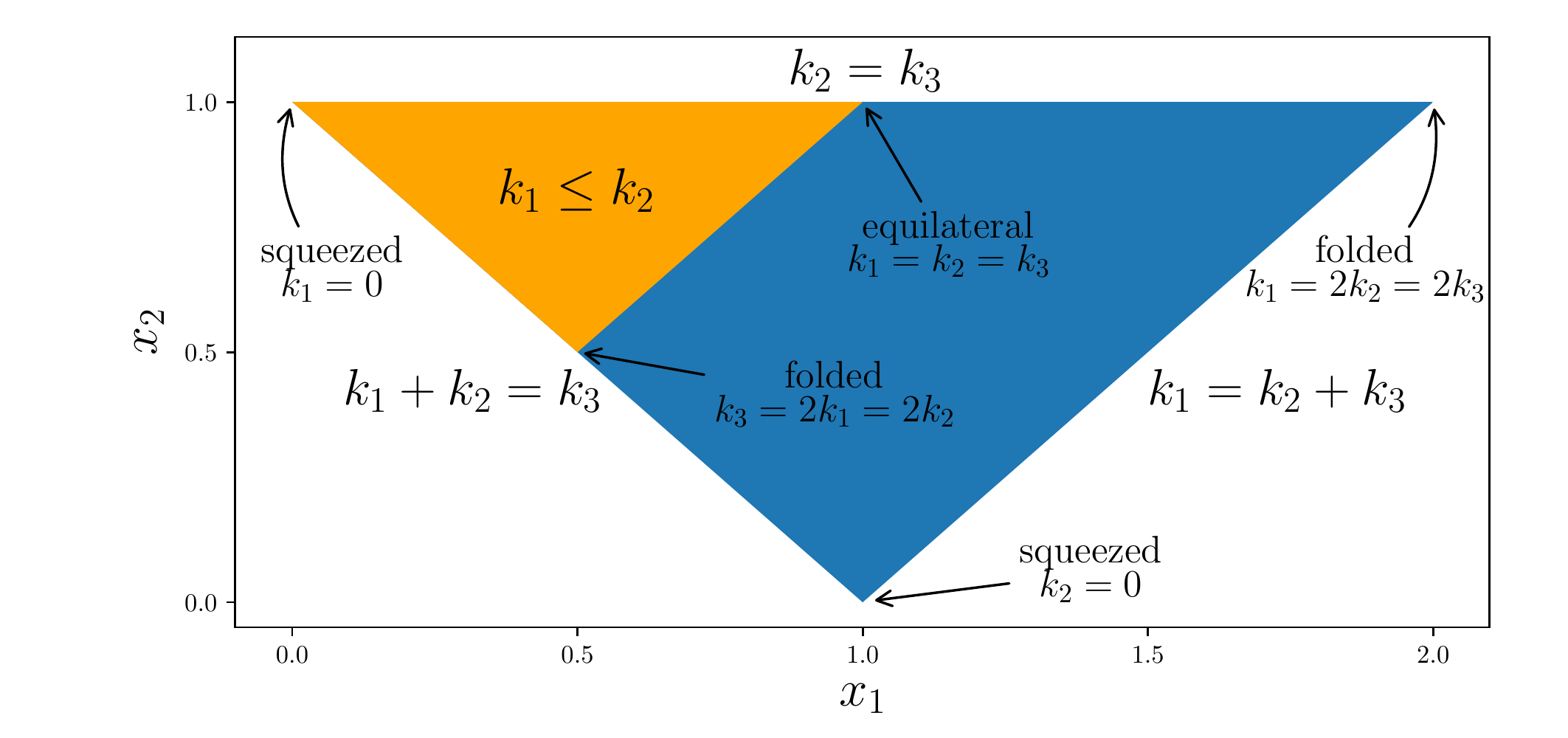}
\caption{Region of the shape of non-gaussianity in the space of $x_1 \equiv k_1 / k_3$ and $x_2 \equiv k_2 / k_3$. Combining triangular inequalities and the condition to avoid redundancy ($k_2 \le k_3$), the colored region (blue$+$orange) exhausts all the configurations of a triangle formed by $\bm{k}_1$, $\bm{k}_2$ and $\bm{k}_3$. In the case of auto-three-point correlations, one can impose a further redundancy condition $k_1 \le k_2$, and the orange shaded region is sufficient, but in our present case, the region has to be extended to the entire colored region.}
\label{fig:shape_region}
\end{figure}
This region is shown in Fig.~\ref{fig:shape_region}. In the case of three-point auto correlations, since $\bm{k}_1$ would also be symmetric together with $\bm{k}_2$ and $\bm{k}_3$, one could further restrict the region by imposing $x_1 \le x_2$, leading to the orange part in Fig.~\ref{fig:shape_region}. However, in the current case for cross correlations, we need to extend the region to include the blue part of the figure.

In order to see different contributions from different terms in $\delta_3 H_{\rm int}^{STT}$ in \eqref{HISTT_Fourier}, we separately compute the terms of the form $\delta\chi TT$ (second line in \eqref{HISTT_Fourier}), $\delta Q TT$ (third line), $U TT$ (fourth line) and the terms originated from integrating out the non-dynamical mode $Y$ (last two lines). We denote the corresponding operators ${\cal F}^{(I)}$, introduced in \eqref{F-def}, by ${\cal F}^{(\chi)}$, ${\cal F}^{(Q)}$, ${\cal F}^{(U)}$ and ${\cal F}^{(I)}_{\rm rest}$, respectively. Note that ${\cal F}^{(I)}_{\rm rest}$ also has terms that contain $\delta\chi$, $\delta Q$ and $U$, denoting ${\cal F}^{(\chi), (Q), (U)}_{\rm rest}$. Their explicit expressions read
\begin{equation}
\begin{aligned}
{\cal F}^{(\chi)} (\bm{k} , \bm{p} , \bm{q} , \tau' ) & = 
a^3 \, \frac{\lambda}{f} H \, e_{ai}^R (\hat{\bm{p}}) \, e_{ai}^{R} (\hat{\bm{q}}) \left[
\frac{m_Q}{2} \, \frac{\dd^{(\bm{k})}}{\dd\tau'}
+ \frac{q}{aH} \left( \frac{\dd^{(\bm{p})}}{\dd\tau'} + aH \right) \right] \; , 
\\ 
{\cal F}^{(Q)} ( \bm{k} , \bm{p} , \bm{q} , \tau')
& = a^4 g H \, e_{ai}^R (\hat{\bm{p}}) \, e_{ai}^{R} (\hat{\bm{q}})  \left(
\xi - \frac{q}{a H} \right)
\; ,
\\
{\cal F}^{(U)} (\bm{k} , \bm{p} , \bm{q} , \tau' ) 
& 
= - a^3 g \, \frac{q_i p_j}{k} \, e^R_{ai}(\hat{\bm{p}}) \, e^{R}_{aj} (\hat{\bm{q}})
\; ,
\end{aligned}
\label{FI-expression}
\end{equation}
and
\begin{equation}
\begin{aligned}
{\cal F}^{(\chi)}_{\rm rest} ( \bm{k} , \bm{p} , \bm{q} , \tau' ) & 
= - a^4 H^2 \, \frac{\lambda}{f} \, m_Q^2 \, 
\frac{\left( p - q \right) e_{ai}^R (\hat{\bm{p}}) \, e_{ai}^R (\hat{\bm{q}})}{k^2 + 2 a^2 H^2 m_Q^2} 
\left( \frac{\dd^{(\bm{p})}}{\dd\tau'} + a H \right) \; ,
\\
{\cal F}^{(Q)}_{\rm rest} ( \bm{k} , \bm{p} , \bm{q} , \tau' ) & 
= a^3 g \, \frac{\left( p - q \right) e_{ai}^R (\hat{\bm{p}}) \, e_{ai}^R (\hat{\bm{q}})}{k^2 + 2 a^2 H^2 m_Q^2} 
\left( \frac{\dd^{(\bm{k})}}{\dd\tau'} + aH \right)
\left( \frac{\dd^{(\bm{p})}}{\dd\tau'} + aH \right) \; ,
\\
{\cal F}^{(U)}_{\rm rest} ( \bm{k} , \bm{p} , \bm{q} , \tau' ) & 
= 2 a^4 H \, \frac{g m_Q}{k} \, \frac{\left( p - q \right) e_{ai}^R (\hat{\bm{p}}) \, e_{ai}^R (\hat{\bm{q}})}{k^2 + 2 a^2 H^2 m_Q^2} 
\left(
- \frac{\dd^{(\bm{k})}}{\dd\tau'} - a H 
+ \frac{\partial_{\tau'} \left( aQ \right)}{a Q}
\right) 
\left( \frac{\dd^{(\bm{p})}}{\dd\tau'} + a H \right) \; ,
\end{aligned}
\label{FrestI-expression}
\end{equation}
where $H = \dot{a} /a$ is the physical Hubble parameter, and time derivatives $\dd^{(\bm{p})} / \dd \tau'$ act only on quantities $S_{IJ} = {\cal R}_{IK} \mathscr{S}_{KJ}$ and $R_{IJ}= \tilde{\mcR}_{IK} \mathscr{T}^R_{KJ}$ that depend on both $\tau'$ and $\bm{p}$ (thus do not act on background quantities directly).
In obtaining these expressions, we have also used $\bm{k} + \bm{p} + \bm{q} = 0$ thanks to the delta function in \eqref{F-def} and the properties of polarization tensors (see below \eqref{SITI-def}).
Then using \eqref{Bzetahh_general}, we can write $B_{\zeta hh}$ as
\begin{equation}
B_{\zeta h h} (k_1 , k_2 , k_3) \,
= \frac{- g \dot{\chi} H^2}{2^2 (2\pi)^{3/2} \Mpl^4} \, 
\frac{\Delta N_{\chi , k_1}}{k_1^2 k_2^2 k_3^2}
\sum_{I= \chi , Q , U} \left[ {\cal J}^{(I)}(k_1 , k_2 , k_3 , \tau) + {\cal J}^{(I)}_{\rm rest}(k_1 , k_2 , k_3 , \tau) \right]
\; ,
\label{B-expression}
\end{equation}
where 
\begin{equation}
\begin{aligned}
{\cal J}^{(I)}(k_1 , k_2 , k_3 , \tau) & \equiv \frac{k_1 k_2 k_3}{g H^3} 
\sum_{J,M,N}
\int^\tau_{-\infty} \dd\tau' 
\left[ {\cal F}^{(I)}(-\bm{k}_1 , -\bm{k}_2 , -\bm{k}_3 , \tau') + {\cal F}^{(I)}(-\bm{k}_1 , -\bm{k}_3 , -\bm{k}_2 , \tau') \right]
\\ & \quad \times 
{\rm Im} \left[ 
\frac{S_{\chi J, k_1}(\tau) \, R_{h M, k_2}(\tau) \, R_{h N, k_3}(\tau) \, 
S_{IJ, k_1}^*(\tau') \, R_{TM, k_2}^*(\tau') \, R_{TN, k_3}^*(\tau')}{(2^3 k_1 k_2 k_3)^{-1} M_p^{-2}} \right] \; ,
\end{aligned}
\label{JI-def}
\end{equation}
with no summation on index $I$, and the same for ${\cal J}_{\rm rest}^{(I)}$ only with ${\cal F}^{(I)}$ replaced by ${\cal F}^{(I)}_{\rm rest}$.
Due to the scaling of $B_{\zeta hh}$ in \eqref{Bzhh_scaleinv}, one can see that ${\cal J}^{(I)}$ and ${\cal J}^{(I)}_{\rm rest}$ are scaling free: ${\cal J}^{(I)} (s k_1 , s k_2 , s k_3) = {\cal J}^{(I)} (k_1 , k_2 , k_3)$, and the same for ${\cal J}^{(I)}_{\rm rest}$, in de Sitter. 
Thus they can be written in terms of $x_1$ and $x_2$, defined in \eqref{x1x2-def}, and are independent of the size of the triangle. Their explicit expressions are summarized in Appendix \ref{appen:explicit}.

Using ${\cal J}^{(I)}$ and ${\cal J}^{(I)}_{\rm rest}$ in \eqref{JI-explicit} and \eqref{JrestI-explicit}, we compute the shape \eqref{shape-def}
\begin{equation}
{\cal S}_{\zeta hh} ( x_1 , x_2 ) = - {\cal N}' \,
\sum_{I= \chi , Q , U} \left[ {\cal J}^{(I)}(x_1 , x_2) + {\cal J}^{(I)}_{\rm rest}(x_1 , x_2) \right] \; ,
\label{shape-expression}
\end{equation}
where ${\cal N}'$ is an arbitrary normalization factor,%
\footnote{Although $\Delta N_{\chi , k_1}$ depends on $k_1$, we absorb this factor in the definition of $\mathcal{N}'$ and exclude it from the definition of shape for simplicity.} 
and the non-linearity parameter \eqref{fNL-def}
\begin{equation}
f_{\zeta hh}^{\rm NL} =
\frac{- 5 \Delta N_{\chi , k_1}}{6 \left( 2 \pi \right)^{4} {\cal P}_\zeta^2} \, 
\frac{g \dot{\chi} H^2}{\Mpl^4} \, 
\frac{x_1 x_2}{x_1^3 + x_2^3 + 1} \, 
\sum_{I= \chi , Q , U} \left[ {\cal J}^{(I)}(x_1 , x_2) + {\cal J}^{(I)}_{\rm rest}(x_1 , x_2) \right] \; .
\label{fNL-expression}
\end{equation}
In computing ${\cal J}^{(I)}$ and ${\cal J}^{(I)}_{\rm rest}$, we solve the matrix-form equations of motion for the scalar sector, \eqref{scalarEoM}, and the ones for the tensor. We then use the rotation matrix ${\cal R}_{IJ}$ in \eqref{scalar-rotation} for scalar and $\tilde{\cal R}_{IJ}$ in \eqref{tensor-matrices} for tensor to obtain $S_{IJ , k}$ and $R_{IJ , k}$, which we need to perform the time integrals in \eqref{JI-def}. 
We neglect slow-roll corrections and treat all the background quantities as constants except for the scale factor $a = -1 / (H\tau)$.
As fiducial values of parameters, we take the following for the purpose of straightforward comparison with Ref.~\cite{Dimastrogiovanni:2018xnn},
\begin{equation}
m_Q = 3.45 \; , \quad
f = 10^{-2} \Mpl \; , \quad
\epsilon_B = 3 \cdot 10^{-5} \; , \quad
\lambda = 1000 \; ,
\label{parametervalues}
\end{equation}
and other parameters are automatically fixed by attractor solutions \eqref{attractor} as $\xi \simeq m_Q + m_Q^{-1} \simeq 3.74$, $\Lambda \simeq 159$ and $W_{\chi\chi} \simeq -41.3 H^2$, with potential form $W(\chi) = \mu^4 \left[ 1 + \cos(\chi /f) \right]$ and initial condition $\chi_0 = 0.9 \cdot \pi /2$.
We are interested in the effect of production of the tensor modes on the bispectrum $B_{\zeta hh}$, which is localized in time around Hubble crossing for each mode. Focusing our interest on this effect and barring other UV and IR behaviors to take over, we restrict the time integrals in ${\cal J}^{(I)}$ and ${\cal J}^{(I)}_{\rm rest}$ to a limited interval of $\sim 7$ e-folds before and after Hubble crossing to correctly include the production effect. Also, in order to deal with fast oscillations in the integrands, we employ the Clenshaw-Curtis Rule for numerical integration using Mathematica.%
\footnote{To cross-check the validity of this numerical method, we have compared numerical results to a crude analytical estimate, and they match each other up to ${\cal O}(1)$ difference, which we nonetheless expect due to low accuracy of our analytical method.}

\begin{figure}
\centering
\includegraphics[width=0.47\textwidth]{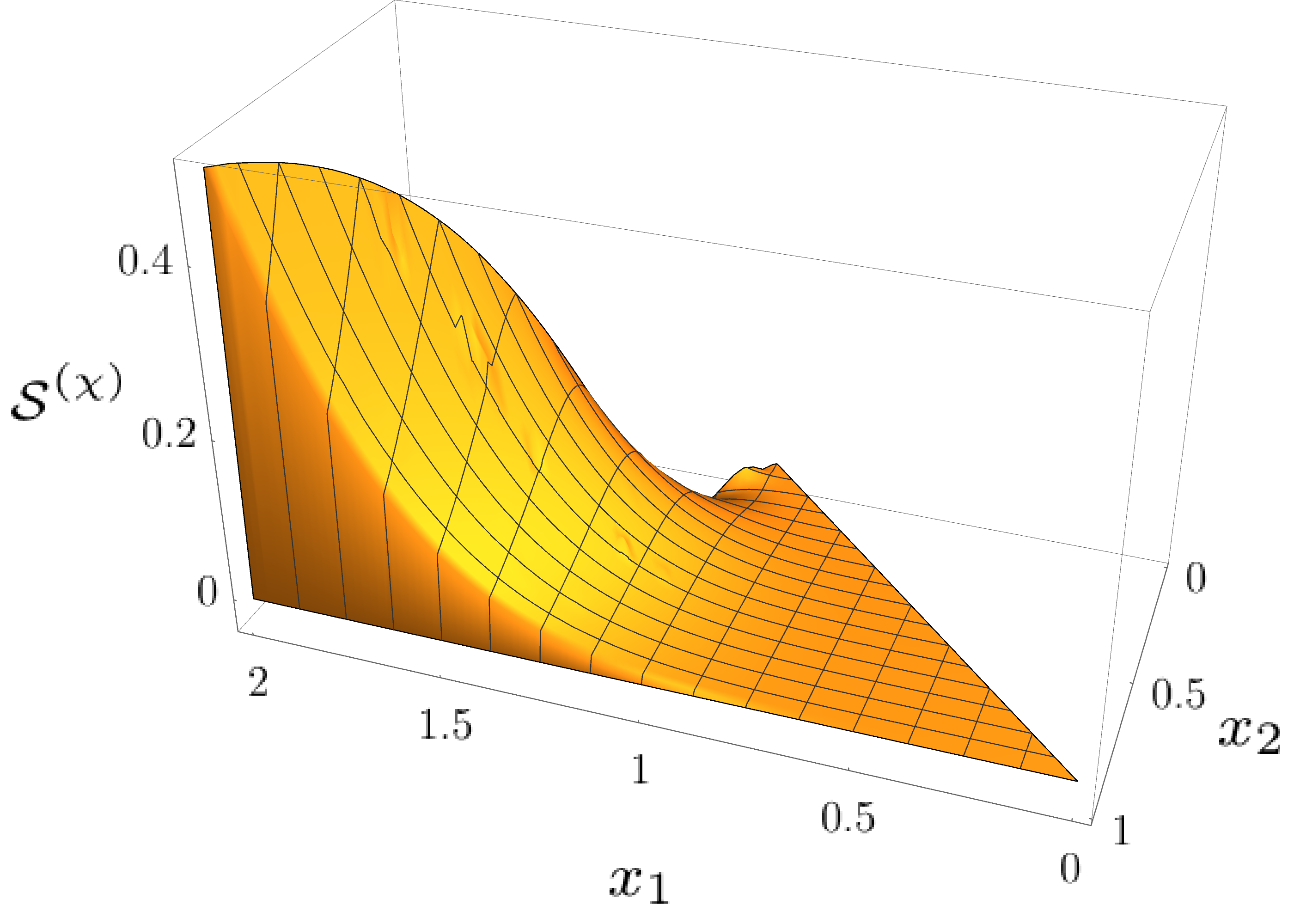} \hfill
\includegraphics[width=0.47\textwidth]{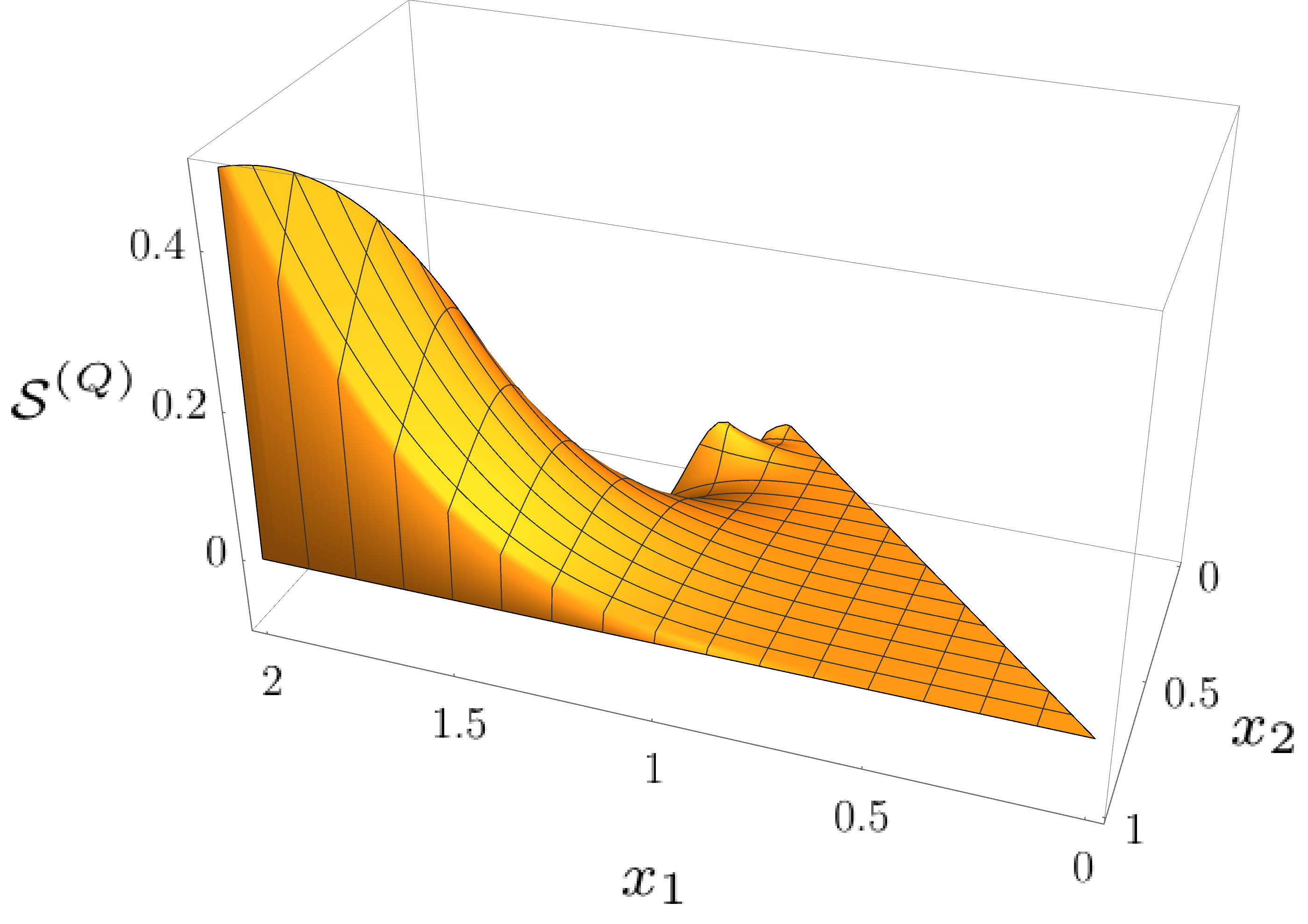}

\vspace{2mm}
\includegraphics[width=0.47\textwidth]{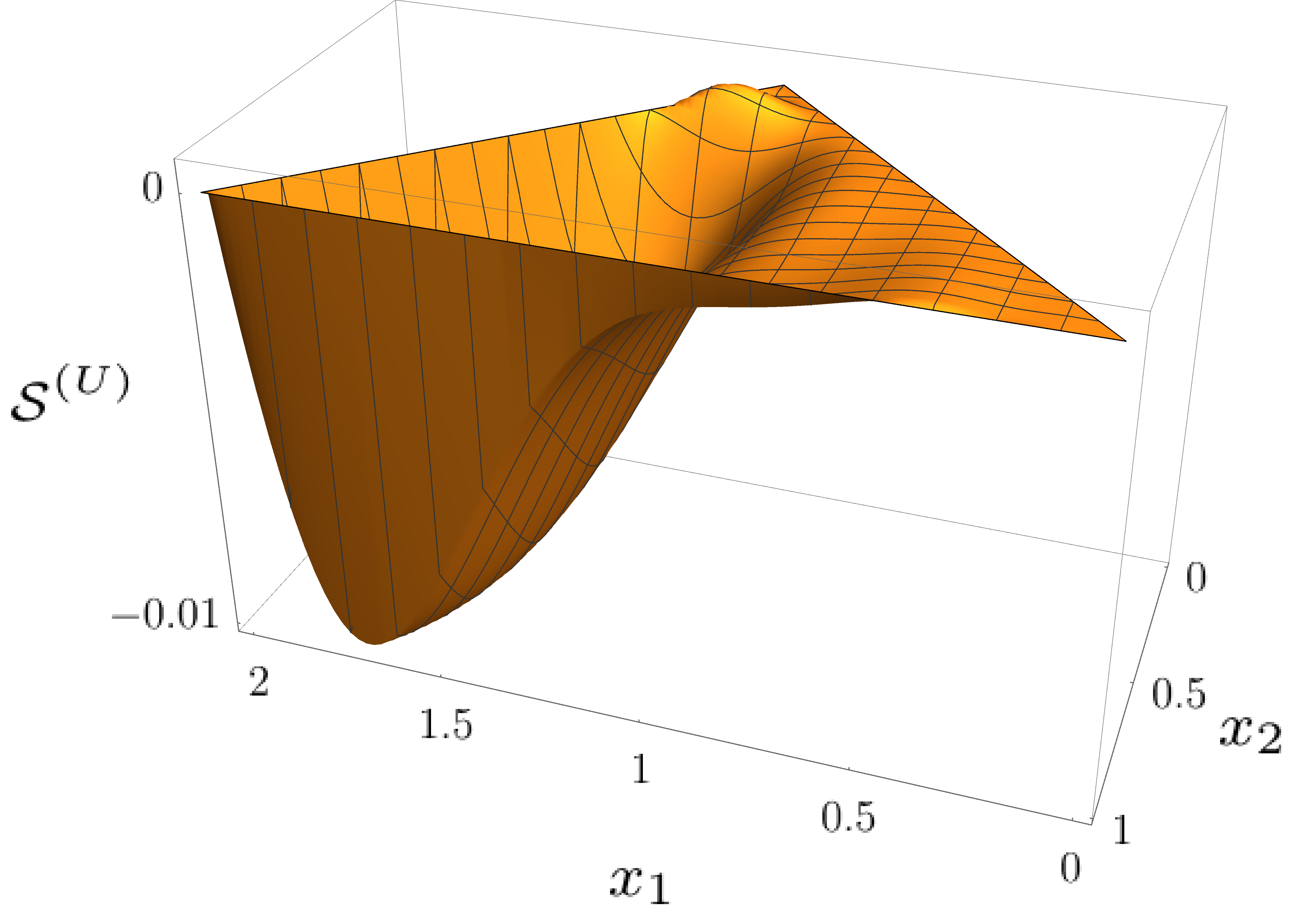} \hfill
\includegraphics[width=0.47\textwidth]{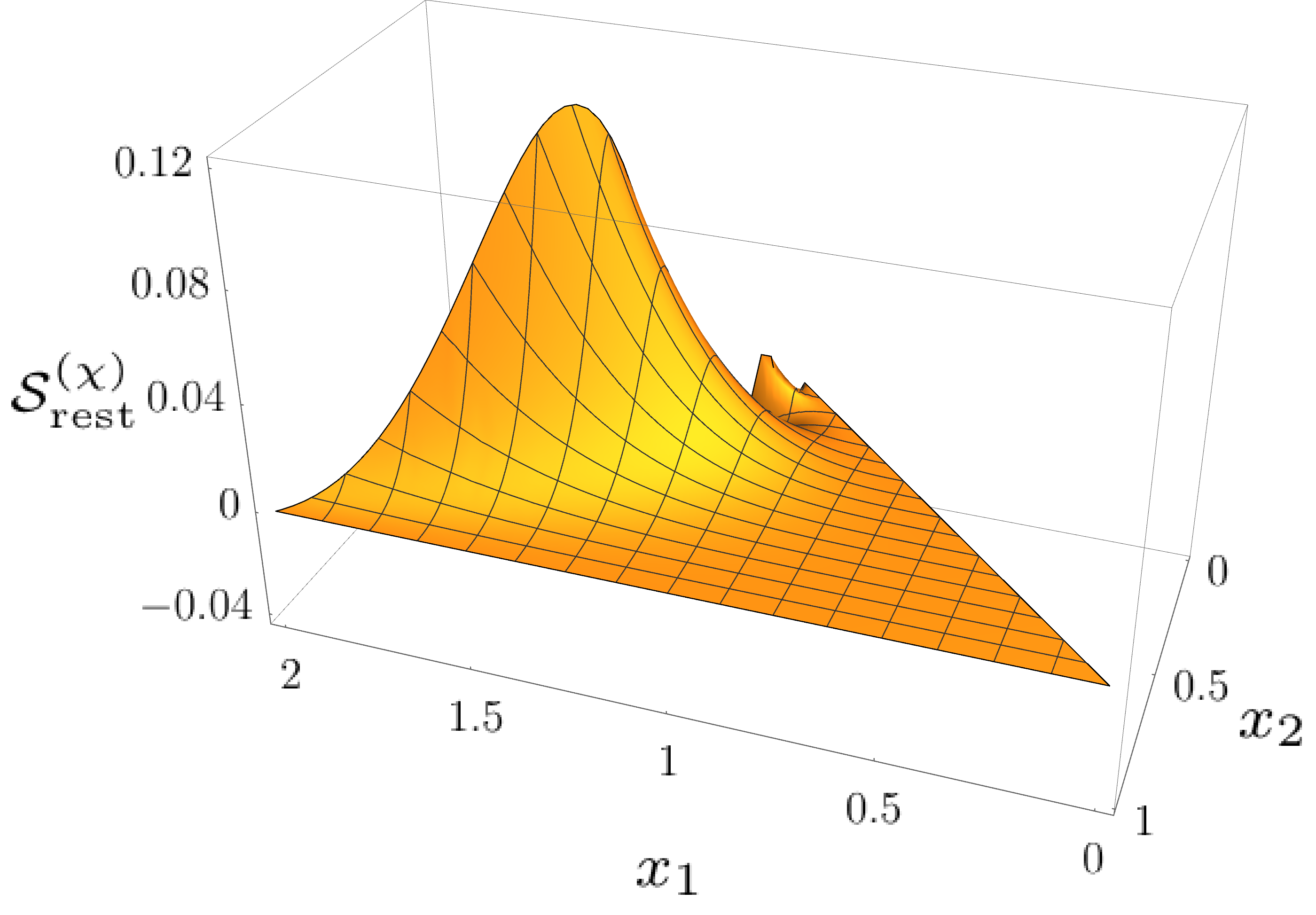}

\vspace{2mm}
\includegraphics[width=0.47\textwidth]{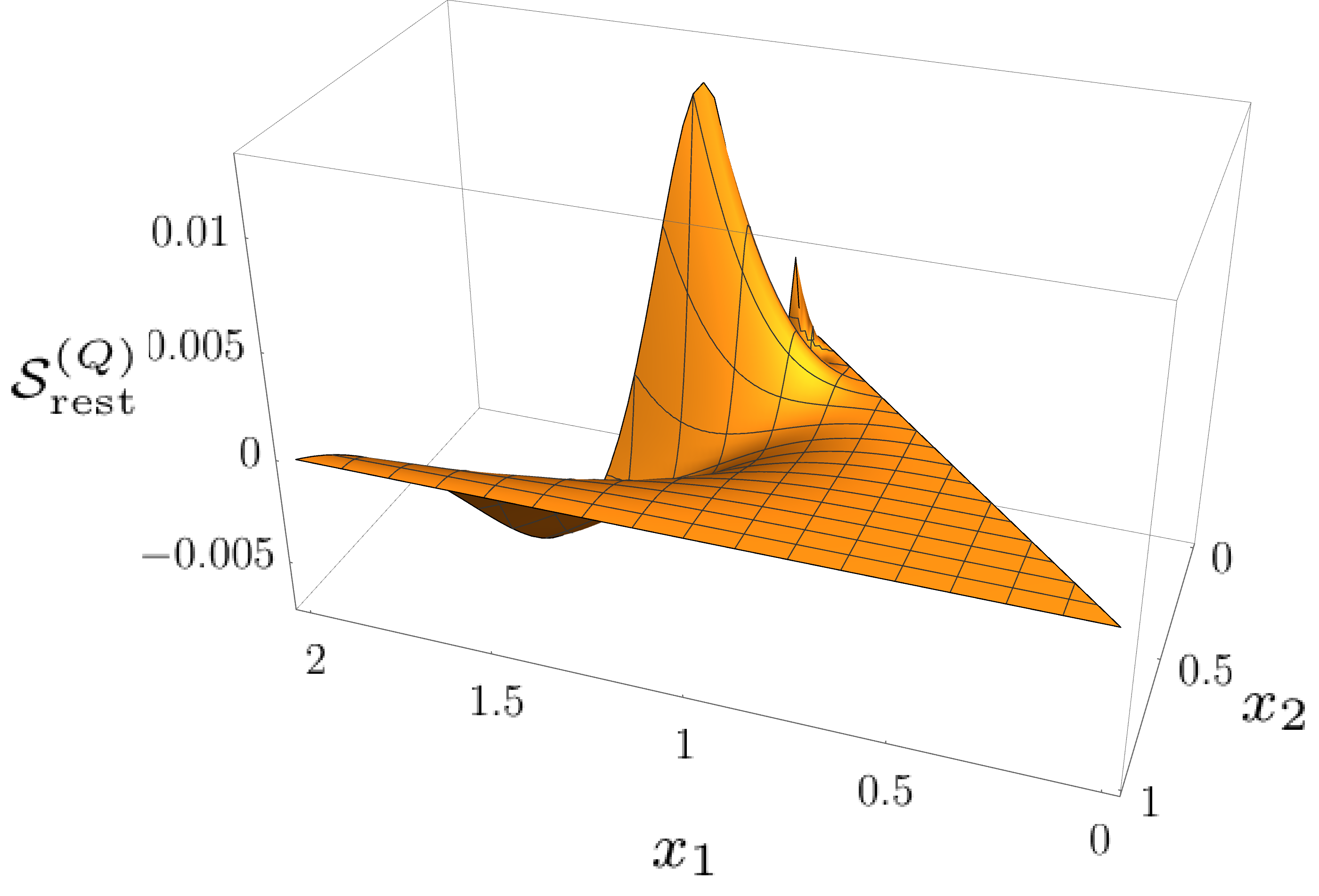} \hfill
\includegraphics[width=0.47\textwidth]{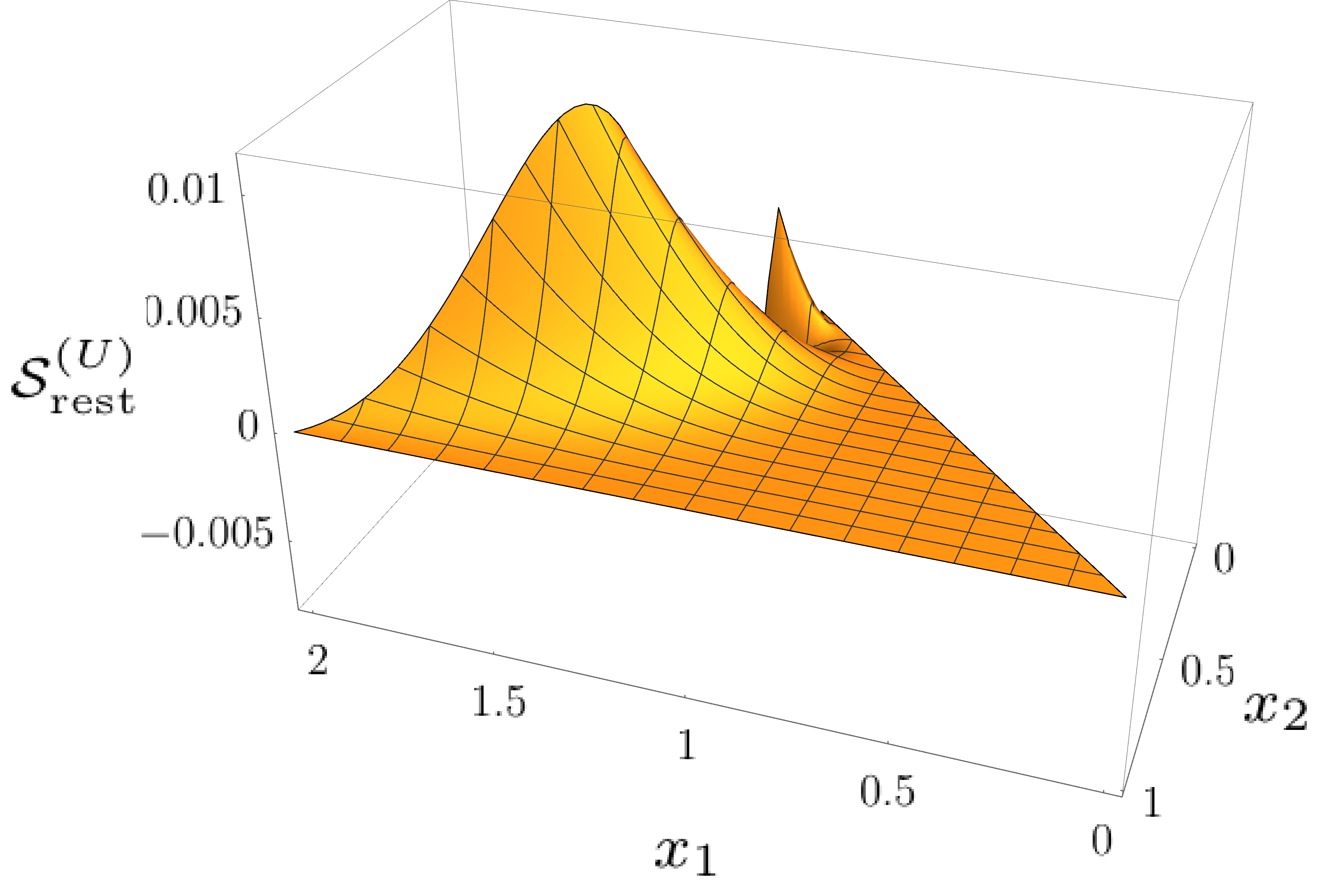}
\caption{The shape of each contribution to ${\cal S}_{\zeta hh}$ in \eqref{shape-expression}. Each of ${\cal S}^{(I)}$ corresponds to the term of ${\cal J}^{(I)}$ respectively, and the same for ${\cal S}^{(I)}_{\rm rest}$. The $z$-axis is normalized such that the sum of all the contributions becomes unity at the folded configuration $x_1 = 2 x_2 = 2$, where the overall signal is peaked.
Note that, for the purpose of presentation, the orientation of the $x$- and $y$-axes is rotated by almost $180$ degrees compared to that in Fig.~\ref{fig:shape_region}.
One can see hierarchical relations $\mcS^{(\chi)}, \mcS^{(Q)}\gtrsim \mcS^{(\chi)}_{\rm rest}\gg \mcS^{(U)}, \mcS^{(Q)}_{\rm rest}, \mcS^{(U)}_{\rm rest}$ in terms of their peak magnitude.}
\label{fig:shape-each}
\end{figure}

\begin{figure}
\centering
\includegraphics[width=0.6\textwidth]{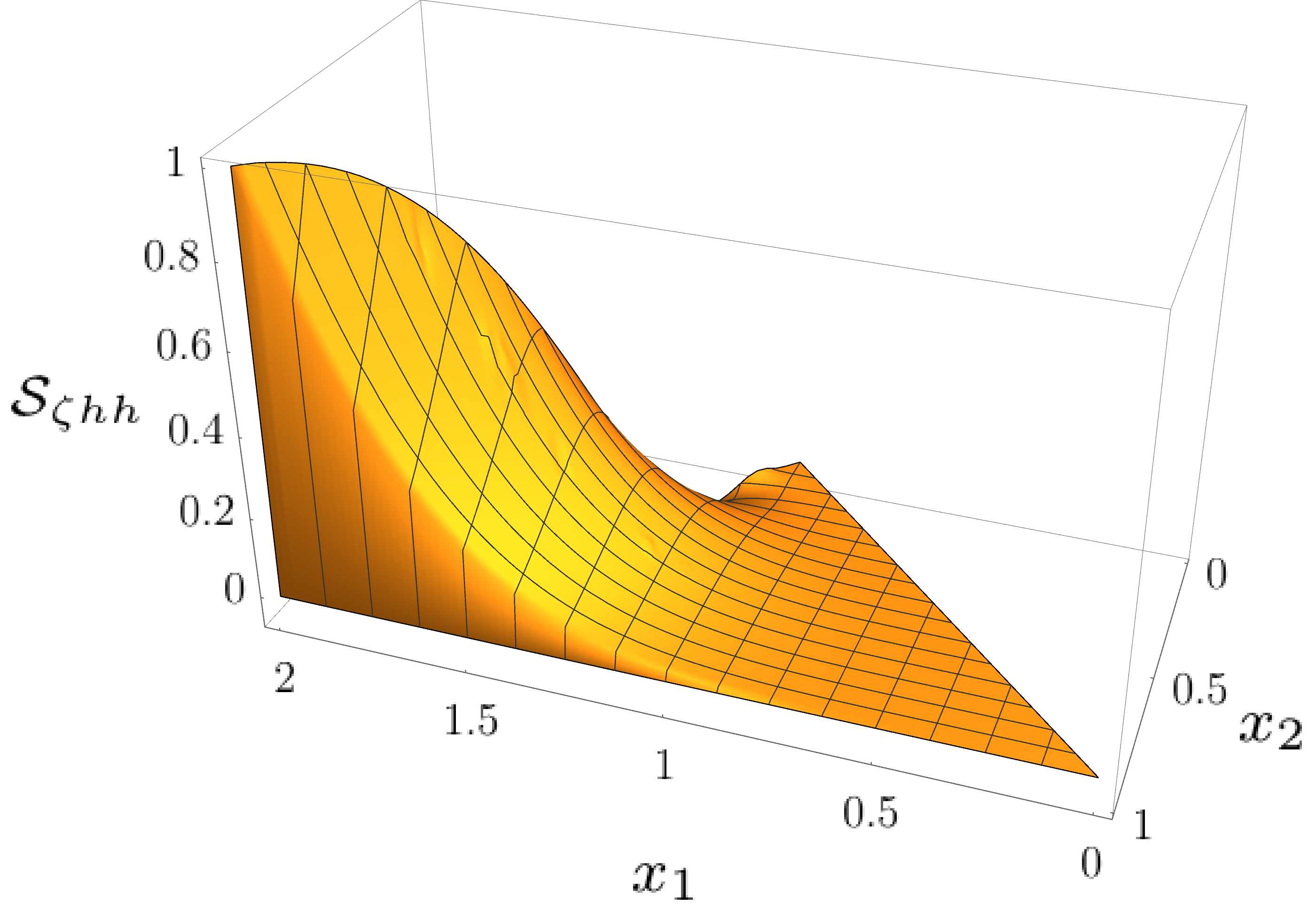} 
\caption{The overall shape of the scalar-tensor-tensor non-gaussianity, \eqref{shape-expression}.
The peak is located at the folded configuration. This is obtained by summing all the contributions plotted in Fig.~\ref{fig:shape-each}.
Note that, as in Fig.~\ref{fig:shape-each}, the orientation of the $x$- and $y$-axes is rotated by almost $180$ degrees compared to that in Fig.~\ref{fig:shape_region}.}
\label{fig:shape-total}
\end{figure}

The result of each contribution to ${\cal S}_{\zeta hh}$ in \eqref{shape-expression} is shown in Fig.~\ref{fig:shape-each}, and the overall shape in Fig.~\ref{fig:shape-total}. The normalization factor ${\cal N}'$ is chosen such that ${\cal S}_{\zeta hh}$ becomes unity at the peak configuration, which is in our case the folded shape, namely ${\cal S}_{\zeta hh} ( x_1 = 2 , x_2 = 1 ) = 1$.
As can be seen both from the figures and from the expressions \eqref{JI-explicit} and \eqref{JrestI-explicit} for ${\cal J}^{(I)}$ and ${\cal J}^{(I)}_{\rm rest}$, every contribution vanishes on the line $x_1 + x_2 - 1 = 0$, i.e.~$k_1 + k_2 = k_3$ (and $1 + x_1 - x_2 = 0$, which is redundant), simply because such an interaction process by one scalar and two tensor modes is not allowed while respecting momentum conservation. On top of that, ${\cal J}^{(U)}$ vanishes at $x_2 + 1 = x_1$, i.e.~$k_1 = k_2 + k_3$, since the vertex for $UTT$ interaction is proportional to $k_{3i} e_{ai}(\hat{\bm{k}}_2)$ (see \eqref{FI-expression}), which must vanish for this configuration as $\hat{\bm{k}}_2 = \hat{\bm{k}}_3$ ($= -\hat{\bm{k}}_1$).
On the other hand, each one of ${\cal J}^{(I)}_{\rm rest}$ vanishes at $k_2 = k_3$ ($x_2 = 1$), as well as $k_1 + k_2 = k_3$. This is because the corresponding part of the interaction Hamiltonian (last two lines of \eqref{HISTT_Fourier}) is antisymmetric in exchange of $\bm{p}$ ($= \bm{k}_2$) and $\bm{q}$ ($= \bm{k}_3$). Due to the nature of the $\langle \zeta h h \rangle$ correlation function, however, $\bm{k}_2$ and $\bm{k}_3$ have to be symmetric, leading to vanishing correlation at $k_2 = k_3$.

From Fig.~\ref{fig:shape-each}, we observe that the dominant contribution comes from ${\cal J}^{(\chi)}$ and ${\cal J}^{(Q)}$, which are similar both in shape and in size. Both peak at the folded configuration $x_1 = 2 x_2 = 2$ (i.e.~$k_1 = 2k_2 = 2k_3$). This can be understood as follows: production of tensor modes is localized in time around Hubble crossing as is seen in Fig.~\ref{PTplot}, and thus correlation between tensor modes is maximum if they cross the horizon at the same time, leading to $k_2 = k_3$. 
On the other hand, the scalar perturbations are more efficiently sourced by the tensor modes in earlier times, because the Green functions of scalar perturbations are decaying around and after the horizon crossing due to their mass.%
\footnote{We call ${\rm Im} \left[ 
S_{\chi J, k_1}(\tau) S_{IJ, k_1}^*(\tau') \right]$ in eq.~\eqref{JI-def} the scalar Green functions. The non-oscillating part of the Green functions start decaying around $k_1 / a \sim 6 H$. This timing is set by the behaviors of eigenvalues of the mass matrix $\Omega^2_{IJ}$ in \eqref{scalar-matrices}.
One of the eigenvalues that corresponds to the lightest eigenmode crosses zero around this time. The timing is independent of the parameters $m_Q$, $\Lambda$ and $W_{\chi\chi}$, as long as $\Lambda \gg m_Q \gg 1$, which is the parameter regime of our interest. The physical interpretation of this specific moment is obscure due to the complicated expressions of the eigenvalues.}
Hence their correlation is maximized when $k_1$ correlates with $k_2$ and $k_3$ modes with hierarchy $k_1 > k_2, k_3$.  Given the momentum conservation $\bm k_1+\bm k_2+\bm k_3=0$, this is achieved at the folded configuration $k_1 = 2 k_2 = 2 k_3$, as we see in our result.
The contributions other than ${\cal J}^{(\chi) , (Q)}$  appear not to follow this argument in Fig.~\ref{fig:shape-each}. In fact, however, the integrals \eqref{JI-explicit} and \eqref{JrestI-explicit} in ${\cal J}^{(I)}$ and ${\cal J}^{(I)}_{\rm rest}$ are maximum at the folded configuration by themselves. Due to the consideration of polarizations discussed in the previous paragraph, these other contributions have prefactors in \eqref{JI-explicit} and \eqref{JrestI-explicit} coming from helicity consideration and have to vanish at the folded configuration $k_1 = 2 k_2 = 2 k_3$.
These contributions are nonetheless subdominant, and the dominant part is controlled by ${\cal S}^{(\chi)}$ and ${\cal S}^{(Q)}$, and therefore the overall shape of non-gaussianity is peaked at the folded configuration, as is seen in Fig.~\ref{fig:shape-total}.

In order to quantify $f_{\zeta hh}^{\rm NL}$, it is more convenient to write \eqref{fNL-expression} in terms of our model parameters. In this regard we replace $\dot\chi_0$, $g$ and $\lambda$ using the relations $\dot\chi_0 = 2fH \xi / \lambda$, $g = m_Q^2 H / (\Mpl \sqrt{\epsilon_B})$ and $\lambda = m_Q \Lambda f / (\Mpl \sqrt{\epsilon_B})$. Moreover, the tensor-to-scalar ratio $r$ in this model can be written as~\cite{Dimastrogiovanni:2016fuu}
\begin{equation}
r = \frac{H^2}{\pi^2 \Mpl^2 {\cal P}_\zeta} \left[ 2 + \epsilon_B {\cal F}^2(m_Q) \right] \; ,
\label{r-expression}
\end{equation}
where the first term in the square parentheses corresponds to the standard prediction from vacuum fluctuations of graviton and the second to the contribution from particle production. The explicit expression of ${\cal F}^2(m_Q)$ can be found in Ref.~\cite{Dimastrogiovanni:2016fuu} and is well fitted for the range $m_Q \in [ 3 , 5]$ by
\begin{equation}
{\cal F}^2 \simeq 0.11 \, m_Q^{7.7} \, {\rm e}^{1.94 m_Q} \; , \qquad
3 \le m_Q \le 5 \; .
\end{equation}
Using \eqref{r-expression} to replace $H/\Mpl$, we can re-express $f_{\zeta hh}^{\rm NL}$ in \eqref{fNL-expression} as
\begin{equation}
f_{\zeta hh}^{\rm NL} =
\frac{- 5 \Delta N_{\chi , k_1}}{2^4 \cdot 3} \, 
\frac{m_Q \xi \, r^2}{\Lambda \left( 2 + \epsilon_B {\cal F}^2 \right)^2} \, 
\frac{x_1 x_2}{x_1^3 + x_2^3 + 1} \, 
\sum_{I= \chi , Q , U} \left[ {\cal J}^{(I)}(x_1 , x_2) + {\cal J}^{(I)}_{\rm rest}(x_1 , x_2) \right] \; ,
\label{fNL-expression2}
\end{equation}
where the dependence on ${\cal P}_\zeta$ is conveniently canceled out.
\begin{figure}
\centering
\includegraphics[width=0.8\textwidth]{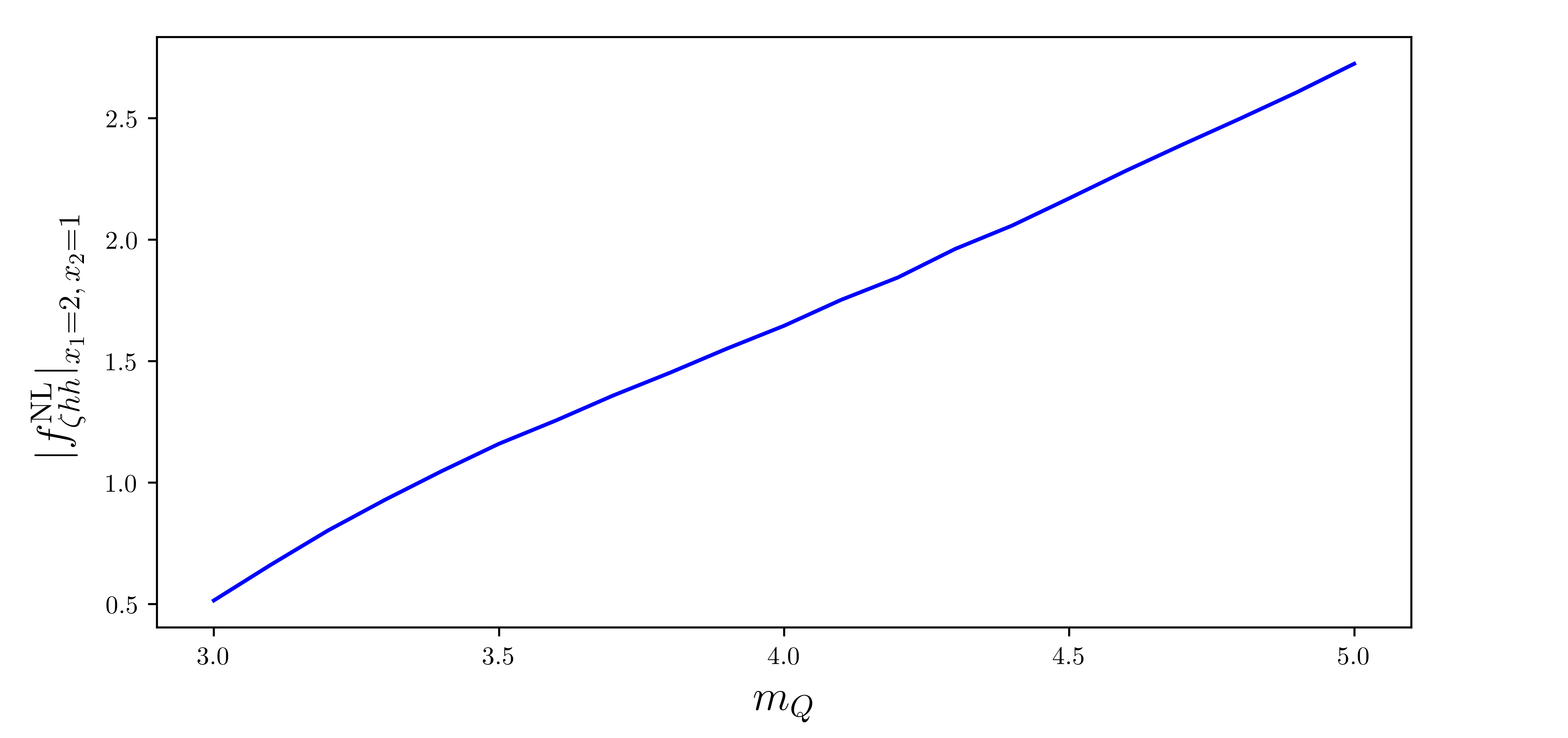}
\caption{The values of $\vert f_{\zeta hh}^{\rm NL} \vert$ at the folded configuration $x_1 = 2 x_2 = 2$ as a function of $m_Q$. 
Here the sign of $f_{\zeta hh}^{\rm NL}$ is in fact negative.
In the numerical computation we set other free parameters as in \eqref{parametervalues}, namely $f = 10^{-2} \Mpl , \, \epsilon_B = 3 \cdot 10^{-5} , \, \lambda = 1000$, and other parameters are automatically fixed. For $f_{\zeta hh}^{\rm NL}$ values, we take $\Delta N_{\chi , k_1} = 60$ and $r = 0.06$.}
\label{fig:fNL_mQ}
\end{figure}
For the fiducial parameters in \eqref{parametervalues}, the value of $f_{\zeta hh}^{\rm NL}$ at the folded configuration $x_1 = 2 x_2 = 2$ is
\begin{equation}
f_{\zeta hh}^{\rm NL} ( x_1 = 2 , \, x_2 = 1 ) \simeq -1.1 \, \frac{\Delta N_{\chi , k_1}}{60} \, \left( \frac{r}{0.06} \right)^2 \; , \qquad
m_Q = 3.45 \; .
\end{equation}
For other values of $m_Q$, we repeat numerical calculations by varying $m_Q$ and plot $f_{\zeta hh}^{\rm NL}$ at the same configuration (which gives the peak amplitudes) for the region $3 \le m_Q \le 5$ in Fig.~\ref{fig:fNL_mQ}. In this plot we fix the model parameters $f = 10^{-2} \Mpl$, $\epsilon_B = 3 \cdot 10^{-5}$ and $\lambda = 1000$ as in \eqref{parametervalues}, and $\Delta N_{\chi , k_1} = 60$ and $r = 0.06$.
We observe that $f_{\zeta hh}^{\rm NL}$ changes almost linearly in $m_Q$; indeed, the exponential factor coming from ${\cal F}^{-4}$ in \eqref{fNL-expression2} is expected to cancel that from ${\cal J}^{(I)}$.
Note that the values of $f_{\zeta hh}^{\rm NL}$ should also depend on other model parameters such as $\Lambda$ and $\epsilon_B$. 
Moreover the total signal of non-gaussianity should receive contributions from the gravitational coupling between $\delta\varphi$ and the scalar modes of $\delta A_\mu^a$ as well, as briefly mentioned at the beginning of Sec.~\ref{sec:cubic}. 
Our main purpose of this study is to report the consistent treatment of mixing effects in cosmological perturbations and implementation of the in-in formalism in this context. We leave the parameter search and detectability against actual observational sensitivities to our future more comprehensive study.

\section{Summary and Discussion}
\label{sec:summary}

In this paper, we considered the coupled system of a spectator axion field and $SU(2)$ gauge fields during inflation.
Axion dynamics in the primordial universe induces several phenomenologically intriguing signatures in the CMB observables. Motion of axion during inflation results in copious production of gauge fields in a parity-violating manner. For Abelian gauge fields, production occurs for each Fourier mode but is localized around its Hubble crossing time. On the other hand, for non-Abelian gauge fields, in particular those of an $SU(2)$ (sub-)group, the degrees of freedom that behave as ``scalar'' under the global part of $SO(3) \cong {\rm adj}[SU(2)]$, which leaves background spatial isotropy intact, can be sustained for a prolonged period during inflation, due to their self interactions as well as their coupling to axion. This in turn leads to an attractor behavior for homogeneous modes, in which the gauge field and axion support each other against decaying away. Inhomogeneous perturbations of these ``scalar'' degrees of freedom inevitably inherit non-trivial behaviors due to the background attractor, and mixings among the ``scalar'' modes are therefore substantial.

Effects of such ``scalar'' mixings are most visible in the mixed scalar-tensor-tensor non-gaussianity $\langle \zeta h h\rangle$ that is sourced by the perturbations of the spectator axion and $SU(2)$ gauge fields. Contrary to the previous works, we treated the mixing of the scalar and tensor fluctuations in a non-perturbative way with the NPS method. As a consistent treatment, we then took into account the contributions of all the relevant interaction vertices, not only $\delta\chi TT$ but also $\delta Q TT$ and $U TT$. We found that the $\delta Q TT$ contribution which had been disregarded as a higher order correction was actually significant. 

We reported our results of $\langle \zeta hh \rangle$ non-gaussianity in Sec.~\ref{sec:results}. Fig.~\ref{fig:shape-each} shows that the signal is dominated by the contributions from the $\delta\chi TT$ and $\delta Q TT$ vertices, which are completely comparable to each other. As clearly seen from Fig.~\ref{fig:shape-total}, the total bispectrum is peaked at the folded configuration in which the wave number of the scalar mode $\zeta$ exceeds the other two of the tensor modes $h$ that are mutually equal, i.e.~$k_1 = 2 k_2 = 2k_3$. This can be understood as follows: the two tensor modes should have the same wave numbers, since the tachyonic enhancement of one of the helicity modes occurs only near Hubble crossing (see Fig.~\ref{PTplot}), and thus their correlation is maximum if they have the same momenta and cross the horizon at the same time. On the other hand, there is no substantial enhancement in the scalar sector; instead, the scalar modes are more efficiently sourced by these tensor modes in earlier time even before the time at $k \sim aH$. 
This is caused by non-trivial mass matrix of the scalar modes $\Omega^2_{IJ}$ in \eqref{scalar-matrices}.
Hence maximal correlation occurs for $k_1 > k_2, k_3$, which, together with the momentum conservation, results in the folded-shape $\langle \zeta hh \rangle$ non-gaussianity. This shape dependence is in large part a result of the scalar mixings and is correctly obtained by the consistent treatment of the mixings. To our knowledge, our result is the first example of folded-shape non-gaussian cross correlations in models of particle production during inflation.
We found that the non-linearity parameter of our $B_{\zeta hh}$ can be ${\cal O}(1)$, as seen in Fig.~\ref{fig:fNL_mQ}. We would like to come back to consideration on the detectability of these parity-violating signals in the CMB observables, such as $TBB$ and $EBB$ correlations, as well as on the comprehensive search for parameter dependence, in our future studies.

Despite of the above interesting results, we do not claim that the computation of $B_{\zeta hh}$ is completed.
In fact, in this paper, we did not calculate all the contributions to the scalar-tensor-tensor non-gaussianity $\langle\zeta h h\rangle$ even at the leading order.  
The scalar perturbations $\delta Q$ and $U$ are gravitationally coupled to the inflaton $\delta\phi$ and thus they induce the curvature perturbation in the same way as $\delta\chi$ does. In these channels, $\langle\delta Q hh\rangle$ and $\langle U hh\rangle$ are proportional to $\langle \zeta hh \rangle$ in similar ways to eq.~\eqref{zetahh-chihh}.
We do not see an obvious reason that these contributions are negligible compared to $\langle\delta \chi hh\rangle$, and there is a chance that they might non-trivially change the result in this paper. 
These contributions can be calculated in essentially the same way as we did in this paper.
We hope to come back to this issue in the near future.

The detectability of the mixed non-gaussianity $\langle \zeta hh\rangle$ has not yet been investigated well.
Although there exist some earlier studies discussing the CMB observations~\cite{Shiraishi:2012sn,Shiraishi:2013vha,Bartolo:2018elp},
those works assumed different production mechanisms of non-gaussianity, the resultant shapes are distinct, and hence they cannot be applied to our case. We need a dedicated work to determine the observability of our result.
We leave this study for future work.

Our calculation method developed in this paper can be applied to other quantities.
It would be interesting to revisit the tensor non-gaussianity $\langle hhh\rangle$ which was computed without the NPS method, while we naively expect that higher order corrections of the mixing terms are not significant in the tensor case as discussed in Sec.~\ref{subsec:outline}.
It is also important to precisely estimate the $1$-loop contribution to the curvature power spectrum $\mcP_\zeta$, because we know the observed value of $\mcP_\zeta$ on the CMB scales to a great precision and hence it potentially puts a tight constraint on this model.
Recently, the $1$-loop $\mcP_\zeta$ was estimated in the model of our interest in \cite{Dimastrogiovanni:2018xnn} and was more thoroughly calculated in the Chromo-natural inflation model in Ref.~\cite{Papageorgiou:2018rfx} in which the mixing effect was treated perturbatively.
We would like to explore them for our future work.

\begin{acknowledgments}
R.N. would like to thank Eamanuela Dimastrogiovanni, Valerie Domcke, Matteo Fasiello, Elisa M.~G.~Ferreira, A.~Emir G\"{u}mr\"{u}k\c{c}\"{u}o\u{g}lu, Kaloian D.~Lozanov, Azadeh Maleknejad, Kyohei Mukaida and Lorenzo Sorbo for helpful discussions.
The work of T.F is partially supported by the Grant-in-Aid for JSPS Research Fellow No.~17J09103.
R.N. was in part supported by the Natural Sciences and Engineering Research Council (NSERC) of Canada
and by the Lorne Trottier Chair in Astrophysics and Cosmology at McGill University.
\end{acknowledgments}

\appendix
\section{Explicit expressions of shape functions}
\label{appen:explicit}

In \eqref{JI-def} we have defined integral expressions $\mathcal{J}^{(I)}$, which in turn characterize the shape of non-gaussianity \eqref{shape-expression}. Their expressions are exact, apart from neglecting the unamplified left-handed modes. In order to perform numerical computations, we take the approximation of background de Sitter, namely constant $H$, $Q$ and $\dot\chi$, as well as $a = - 1 / (H\tau)$.
It is then convenient to define dimensionless variables
\begin{equation}
\tilde{S}_{IJ} (- k \tau) \equiv \frac{k \sqrt{2k}}{H} \, S_{IJ , k}(\tau) \; , \;
\tilde{R}_{hJ}(- k \tau) \equiv \Mpl \, \frac{k \sqrt{2k}}{H} \, R_{hJ , k}(\tau) \; , \;
\tilde{R}_{TJ}(- k \tau) \equiv \frac{k \sqrt{2k}}{H} \, R_{TJ , k}(\tau) \; ,
\end{equation}
where one finds that the tilded variables on the left-hand sides are functions only of $-k\tau$ in the de Sitter limit.
Contractions of polarization tensors in \eqref{FI-expression} and \eqref{FrestI-expression} are found as
\begin{equation}
\begin{aligned}
e^R_{ai}(-\hat{\bm{k}}_2) \, e^R_{ai}(-\hat{\bm{k}}_3) & 
= \frac{1}{4} \left( 1 - \hat{\bm{k}}_2 \cdot \hat{\bm{k}}_3 \right)^2 
= \frac{\left( 1 + x_1 - x_2 \right)^2 \left( x_1 + x_2 - 1 \right)^2}{16 x_2^2}
\; ,
\\
\hat{\bm{k}}_{3 i} \hat{\bm{k}}_{2 j} \, e^R_{ai}(-\hat{\bm{k}}_2) \, e^R_{aj}(-\hat{\bm{k}}_3) & 
= \frac{1}{4} \left( 1 - \hat{\bm{k}}_2 \cdot \hat{\bm{k}}_3 \right)
\left[ 1 - \left( \hat{\bm{k}}_2 \cdot \hat{\bm{k}}_3 \right)^2 \right]
\\ &
= \frac{\left( 1 + x_1 - x_2 \right)^2 \left( x_1 + x_2 - 1 \right)^2 \left( x_2 + 1 - x_1 \right) \left( x_1 + x_2 + 1 \right)}{32 x_2^3} \; .
\end{aligned}
\end{equation}
Using these, some algebra arrives at the explicit forms of ${\cal J}^{(I)}$ given by
\begin{equation}
\begin{aligned}
{\cal J}^{(\chi)}(x_1 , x_2 , \tau) & =
\frac{\lambda H}{g f}
\frac{\left( 1 + x_1 - x_2 \right)^2 \left( x_1 + x_2 - 1 \right)^2}{16 \, x_1 x_2^3}
\sum_{I,J,K}
\\ & \quad \times 
\int_z^{\infty} \frac{\dd z'}{z'^3}
\left[ 
- x_1 m_Q \, \frac{\dd^{(k_1)}}{\dd ( x_1 z')}
- x_2 z' \left( \frac{\dd^{(k_2)}}{\dd (x_2 z')} + \frac{\dd^{(k_3)}}{\dd z'} \right)
+ 1 + x_2 \right]
\\ & \qquad\quad\; \times 
{\rm Im} \left[ 
\tilde S_{\chi I}(x_1 z) \, \tilde R_{h J}(x_2 z) \, \tilde R_{h K}(z) \, 
\tilde S_{\chi I}^*(x_1 z') \, \tilde R_{TJ}^*(x_2 z') \, \tilde R_{TK}^*(z') \right] \; ,
\\
{\cal J}^{(Q)}(x_1 , x_2 , \tau) & =
\frac{\left( 1 + x_1 - x_2 \right)^2 \left( x_1 + x_2 - 1 \right)^2}{16 \, x_1 x_2^3}
\sum_{I,J,K}
\int_z^{\infty} \frac{\dd z'}{z'^4}
\left[ 
2 \xi - \left( 1+ x_2 \right) z' \right]
\\ & \qquad\quad\; \times 
{\rm Im} \left[ 
\tilde S_{\chi I}(x_1 z) \, \tilde R_{hJ}(x_2 z) \, \tilde R_{hK}(z) \, 
\tilde S_{QI}^*(x_1 z') \, \tilde R_{TJ}^*(x_2 z') \, \tilde R_{TK}^*(z') \right] \; ,
\\
{\cal J}^{(U)}(x_1 , x_2 , \tau) & = -
\frac{\left( 1 + x_1 - x_2 \right)^2 \left( x_1 + x_2 - 1 \right)^2 \left( x_2 + 1 - x_1 \right) \left( x_1 + x_2 + 1 \right)}{16 \, x_1^2 x_2^3} 
\sum_{I,J,K}
\\ & \quad \times 
\int_z^{\infty} \frac{\dd z'}{z'^3} \,
{\rm Im} \left[ 
\tilde S_{\chi I}(x_1 z) \, \tilde R_{hJ}(x_2 z) \, \tilde R_{hK}(z) \, 
\tilde S_{UI}^*(x_1 z') \, \tilde R_{TJ}^*(x_2 z') \, \tilde R_{TK}^*(z') \right] \; ,
\end{aligned}
\label{JI-explicit}
\end{equation}
and
\begin{equation}
\begin{aligned}
{\cal J}^{(\chi)}_{\rm rest}(x_1 , x_2 , \tau) & =
- \frac{\lambda H m_Q^2}{g f} \,
\frac{\left( 1 + x_1 - x_2 \right)^2 \left( x_1 + x_2 - 1 \right)^2 \left( 1 - x_2 \right)}{16 \, x_1 x_2^3}
\sum_{I,J,K}
\\ & \quad \times
\int_z^{\infty} \frac{\dd z'}{z'^2} \,
\frac{1}{x_1^2 z'^2 + 2 m_Q^2}
\left[ x_2 \frac{\dd^{(k_2)}}{\dd (x_2 z')} - \frac{\dd^{(k_3)}}{\dd z'} \right]
\\ & \qquad\quad\; \times 
{\rm Im} \left[ 
\tilde S_{\chi I}(x_1 z) \, \tilde R_{hJ}(x_2 z) \, \tilde R_{hK}(z) \, 
\tilde S_{\chi I}^*(x_1 z') \, \tilde R_{TJ}^*(x_2 z') \, \tilde R_{TK}^*(z') \right] \; ,
\\
{\cal J}^{(Q)}_{\rm rest}(x_1 , x_2 , \tau) & = -
\frac{\left( 1 + x_1 - x_2 \right)^2 \left( x_1 + x_2 - 1 \right)^2 \left( 1 - x_2 \right)}{16 \, x_1 x_2^3}
\sum_{I,J,K}
\\ & \quad \times
\int_z^{\infty} \frac{\dd z'}{z'^2} \,
\frac{1}{x_1^2 z'^2 + 2 m_Q^2} \,
\left( x_1 z' \, \frac{\dd^{(k_1)}}{\dd (x_1 z')} - 1 \right) \left( x_2 \, \frac{\dd^{(k_2)}}{\dd (x_2 z')} - \frac{\dd^{(k_3)}}{\dd z'} \right) 
\\ & \qquad\quad\; \times 
{\rm Im} \left[ 
\tilde S_{\chi I}(x_1 z) \, \tilde R_{hJ}(x_2 z) \, \tilde R_{hK}(z) \, 
\tilde S_{QI}^*(x_1 z') \, \tilde R_{TJ}^*(x_2 z') \, \tilde R_{TK}^*(z') \right] \; ,
\\
{\cal J}^{(U)}_{\rm rest}(x_1 , x_2 , \tau) & = m_Q \,
\frac{\left( 1 + x_1 - x_2 \right)^2 \left( x_1 + x_2 - 1 \right)^2 \left( 1 - x_2 \right)}{8 \, x_1 x_2^3}
\sum_{I,J,K}
\\ & \quad \times
\int_z^{\infty} \frac{\dd z'}{z'^2} \,
\frac{1}{x_1^2 z'^2 + 2 m_Q^2} \,
\frac{\dd^{(k_1)}}{\dd (x_1 z')} 
\left( x_2 \, \frac{\dd^{(k_2)}}{\dd (x_2 z')} - \frac{\dd^{(k_3)}}{\dd z'} \right)
\\ & \qquad\quad\; \times 
{\rm Im} \left[ 
\tilde S_{\chi I}(x_1 z) \, \tilde R_{hJ}(x_2 z) \, \tilde R_{hK}(z) \, 
\tilde S_{UI}^*(x_1 z') \, \tilde R_{TJ}^*(x_2 z') \, \tilde R_{TK}^*(z') \right] \; ,
\end{aligned}
\label{JrestI-explicit}
\end{equation}
where $z \equiv - k_3 \tau$ and $z' \equiv - k_3 \tau'$.
We use these expressions of ${\cal J}^{(I)}$ and ${\cal J}^{(I)}_{\rm rest}$ for numerical integrations.
As described below \eqref{parametervalues}, we restrict the domain of integrals around horizon crossing, in order to correctly take into account the effects from production of the gauge-field tensor mode. Under this restriction, the result is independent of time $\tau$. For concreteness, we take $0.03 \le z \le 20$ as the domain of integration.

\section{Comparison to the previous work}
\label{appen:comparison}

In this appendix, we compare our work with the previous work~\cite{Dimastrogiovanni:2018xnn}.
Although the same bispectrum $B_{\zeta h h}$ as ours was calculated based on the same model, the following two major differences between our approach and theirs are found: (i) In Ref.~\cite{Dimastrogiovanni:2018xnn}, the mixing terms between the scalar perturbations were ignored, and $\delta\chi$ sourced by $(T^R)^2$ is computed with the Green function of a massless scalar field. However, taking into account the mixing terms with the NPS formalism, we have calculated the sourced $\delta\chi$ in a more accurate manner (see Fig.~\ref{Pchi-comparison}).
(ii) A different relation between $\delta\chi$ and $\zeta$ was considered in Ref.~\cite{Dimastrogiovanni:2018xnn}. While we have used eq.~\eqref{sourced_zeta} in which $\delta\chi$ induces $\delta\phi$, which is connected to $\zeta$ in the standard way \eqref{zeta-def}, the authors in Ref.~\cite{Dimastrogiovanni:2018xnn} considered a different channel where $\delta\chi$ directly contributes to $\zeta$ without the aid of $\delta\phi$.

We discuss the consequences of the above differences, (i) and (ii).
Ref.~\cite{Dimastrogiovanni:2018xnn} argued that the bispectrum $B_{\zeta hh}$ has a peak at the equilateral configuration. However, we have found that the peak is located at the folded configuration.
This discrepancy must originate from the first difference (i) in the treatment of the mixing terms, because the linear relation between $\delta \chi$ and $\zeta$ does not change the bispectrum shape.
Then, the consequence of the second difference (ii) is of interest. 
Actually, the difference (ii) leads to only an overall factor of the bispectrum, as we will see below.

The curvature perturbation on the flat gauge is generally given by
\begin{equation}
\zeta(t,\bm x)=-H(t)\frac{\delta\rho(t,\bm x)}{\dot{\rho}_0(t)},
\end{equation}
where $\rho_0$ is the total energy density at the background and $\delta\rho$ is its perturbation. Focusing on the inflaton and the spectator axion, and considering the slow-roll regime of $\chi_0$ during inflation, one finds
\begin{equation}
\zeta\supset -H\frac{\delta\rho_\phi+\delta\rho_\chi}{\dot{V}(\phi_0)}
\simeq-\frac{H}{\dot{\phi}_0}\,\delta\varphi-\frac{HW_\chi}{V_\phi \dot{\phi}}\,\delta\chi\,,
\label{total zeta}
\end{equation}
where the kinetic energies of the fields are ignored and the background energy density is assumed to be dominated by the inflaton potential, $\rho_0\simeq V(\phi_0)$.
Although our discussion around eq.~\eqref{sourced_zeta} has only considered the first term in eq.~\eqref{total zeta}, the axion perturbation $\delta\chi$ should also contribute to $\zeta$ through the second term as well, which was the one computed in Ref.~\cite{Dimastrogiovanni:2018xnn}.

Let us compare the two terms in eq.~\eqref{total zeta}.
The $\delta\chi$ contribution through the first term in eq.~\eqref{total zeta} is given in eq.~\eqref{sourced_zeta}. On the other hand, the second term can be rewritten as
\begin{equation}
\hat\zeta^{(\chi)}  \simeq - \frac{H W_\chi }{V_\phi \dot{\phi}_0}\,\delta\hat{\chi}
\simeq -\frac{\lambda \epsilon_B}{2f m_Q \epsilon_\phi}\delta\hat{\chi}\,,
\label{zeta chi}
\end{equation}
where $3H\dot{\phi}_0+V_\phi\simeq 0$, $\dot{\phi}_0^2 = 2\epsilon_\phi H^2\Mpl^2$ and \eqref{attractor} are used, and $\hat{\zeta}^{(\chi)}$ denotes the direct contribution from the energy density of the axion perturbation $\delta\rho_\chi$ to the curvature perturbation.
Plugging our parameters, one finds that the ratio of the two terms is
\begin{equation}
\frac{\hat{\zeta}^{(\chi)}}{\hat{\zeta}^{(s)}}
\simeq \frac{\lambda^2}{4(m_Q^2+1)\Delta N_{\chi,k}}
\frac{\epsilon_B \Mpl^2}{\epsilon_\phi f^2}
\approx 3.2\times 10^4 \left(\frac{\epsilon_\phi}{3\times 10^{-3}}\right)^{-1}
\left(\frac{\Delta N_{\chi,k}}{60}\right)^{-1},
\label{zeta ratio}
\end{equation}
where $\epsilon_\phi=3\times 10^{-3}$ was used in Ref.~\cite{Dimastrogiovanni:2018xnn}.
Thus, it appears that the direct contribution from $\delta\chi$ to $\zeta$ is dominant compared to the channel through $\delta\varphi$, and the leading-order bispectrum can be obtained by simply multiplying our result eq.~\eqref{B-expression} by this ratio eq.~\eqref{zeta ratio} in that case.
Nevertheless, that is not necessarily correct, since the above comparison applies only during the period when the slow-roll approximations, both inflationary and for the axion-gauge field dynamics, are valid, and the curvature perturbation observed at the late time can be substantially different from $\zeta^{(\chi)}$.

It is well known that the curvature perturbation is conserved on super-horizon scales, if isocurvature perturbation is absent.
In the current case, however, we have isocurvature perturbations, and hence the curvature perturbation is not necessarily conserved. Indeed, $\zeta^{(\chi)} \propto \delta\chi$ rapidly decays, once $\chi_0$ ceases the slow roll by approaching to its potential minimum and starts a damped oscillation during or soon after inflation.
In that case, $\delta\rho_\chi$ becomes negligible and  $\zeta^{(\chi)}$ virtually vanishes. 
On the other hand, $\delta_m \varphi$ in eq.~\eqref{sourced_zeta} which has been sourced by $\delta\chi$ does not disappear, even if the source term $\delta\chi$ vanishes.
Therefore, it is conservative to estimate the bispectrum $B_{\zeta hh}$ by considering only the adiabatic part of the contribution from $\delta\chi$ to $\zeta$ during inflation, as we have done.

$\zeta^{(\chi)}$ can remain relevant, if the spectator axion occupies a significant fraction of the total energy density after inflation and the curvaton-like mechanism works ~\cite{Enqvist:2001zp,Lyth:2001nq,Moroi:2002rd}.
Thus, the result of $B_{\zeta hh}$ highly depends on the dynamics of the inflaton and the spectator axion after inflation.
For instance, provided that the spectator axion $\chi$ oscillates around a quadratic potential after the inflaton decays and $\chi$ subsequently decays, its contribution to the curvature perturbation is given by~\cite{Sasaki:2006kq,Enqvist:2013paa}
\begin{equation}
\hat{\zeta}^{(\chi)}=\frac{2\rho_\chi}{4\rho_{\rm rad}+3\rho_\chi}\frac{\delta\hat{\chi}}{\chi_0},
\label{curvaton eq}
\end{equation}
where $\rho_{\rm rad}\propto a^{-4}$ is the radiation energy density
and the sudden decay approximation is used.
This equation should be evaluated when $\chi$ decays, because $\hat{\zeta}^{(\chi)}$
is frozen after that.
In this case, eq.~\eqref{zeta chi} should be replaced by eq.~\eqref{curvaton eq} and then the significance of the $\delta\chi$ contribution may substantially change from eq.~\eqref{zeta ratio}.
Furthermore, one should check if the direct contribution from $\delta\chi$ to $\zeta$ does not ruin the observed curvature power spectrum $\mcP_\zeta$ and satisfies the constraint on the bispectrum $B_{\zeta}$ for consistency.
In particular, since $\delta\hat{\chi}$, which is sourced by $(T^R)^2$, is totally non-gaussian, the observational bound on the scalar non-gaussianity may restrict such possibilities.

\end{document}